\documentclass[useAMS,usenatbib,twocolumn]{mn2e}
\usepackage{graphicx}
\usepackage{mn2e-breakabs}
\begin{document}
 
\title[Ameliorating Systematic Uncertainties in Clustering Measurements]{Ameliorating Systematic Uncertainties in the Angular Clustering of Galaxies: A Study using SDSS-III}

\author[A. J. Ross et al.]{\parbox{\textwidth}{Ashley J. Ross\thanks{Email: Ashley.Ross@port.ac.uk}$^{1}$, Shirley Ho$^{2}$, 
Antonio J. Cuesta$^{3}$, 
Rita Tojeiro$^{1}$,
 Will J. Percival$^{1}$, 
 David Wake$^3$, 
 Karen L. Masters$^{1,4}$, 
 Robert C. Nichol$^{1,4}$, 
 Adam D. Myers$^{5}$,
 Fernando de Simoni$^{6,7}$, 
 Hee Jong Seo$^{8}$, 
 Carlos Hern\'andez-Monteagudo$^{9,10}$, 
 Robert Crittenden$^{1}$, 
 Michael Blanton$^{11}$, 
 J. Brinkmann$^{12}$, 
  Luiz A. N. da Costa$^{6,7}$, 
  Hong Guo$^{13}$, 
  Eyal Kazin$^{11}$, 
  Marcio A. G. Maia$^{6,7}$, 
  Claudia Maraston$^{1}$, 
  Nikhil Padmanabhan$^{3}$, 
  Francisco Prada$^{14}$, 
  Beatriz Ramos$^{6,7}$, 
  Ariel Sanchez$^{15}$, 
  Edward F. Schlafly$^{16}$, 
  David J. Schlegel$^2$, 
  Donald P. Schneider$^{17}$, 
  Ramin Skibba$^{18}$, 
  Daniel Thomas$^{1,4}$, 
  Benjamin A. Weaver$^{11}$, 
  Martin White$^{2,19,20}$, 
  Idit Zehavi$^{13}$}
\vspace*{4pt} \\
$^1$Institute of Cosmology \& Gravitation, Dennis Sciama Building, University of Portsmouth, Portsmouth, PO1 3FX, UK\\
$^2$Lawrence Berkeley National Laboratory, 1 Cyclotron Road, Berkeley, CA 94720, USA\\
$^3$Yale Center for Astronomy and Astrophysics, Yale University, New Haven, CT 06511, USA\\
$^4$SEPnet, South East Physics Network (www.sepnet.ac.uk)\\
$^5$Department of Physics and Astronomy, University of Wyoming, Laramie, WY 82071, USA\\
$^6$Observat\'orio Nacional, Rua Gal. Jos\'e Cristino 77, Rio de Janeiro, RJ - 20921-400, Brazil\\
$^7$Laborat\'orio Interinstitucional de e-Astronomia - LineA, Rua Gal. Jos\'e Cristino 77, Rio de Janeiro, RJ - 20921-400, Brazil\\
$^8$Berkeley Center for Cosmological Physics, LBL and Department of Physics, University of California, Berkeley, CA, USA 94720\\
$^9$Max Planck-Institut f\"ur Astrophysik (MPA), Karl Schwarzschild Str.1,
Garching bei M\"unchen, D-85741, Germany\\
$^{10}$Centro de Estudios de F\' \i sica del Cosmos de Arag\'on (CEFCA), Plaza
San Juan, 1, Planta 2, E-44001, Teruel, Spain\\
$^{11}$Center for Cosmology and Particle Physics, New York University, New York, NY 10003, USA\\
$^{12}$Apache Point Observatory, 2001 Apache Point Road, Sunspot, NM 88349, USA\\
$^{13}$Department of Astronomy, Case Western Reserve University, Cleveland, Ohio 44106, USA\\
$^{14}$Instituto de Astrofisica de Andalucia (CSIC), Granada, Spain\\
$^{15}$Max-Planck-Institut f\"ur extraterrestrische Physik, Giessenbachstr. 1,
85748 Garching, Germany\\
$^{16}$Department of Physics, Harvard University, 17 Oxford Street, Cambridge, MA 02138, USA\\
$^{17}$Department of Astronomy \& Astrophysics, The Pennsylvania State University, 525 Davey Lab, University Park, PA 16802, USA\\
$^{18}$Steward Observatory, University of Arizona, 933 North Cherry Ave., Tucson, AZ 85721, USA\\
$^{19}$Department of Physics, University of California, 366 LeConte Hall, Berkeley, CA 94720, USA\\
$^{20}$Department of Astronomy, 601 Campbell Hall, University of California at Berkeley, Berkeley, CA 94720, USA
}

\date{Accepted by MNRAS}

\pagerange{\pageref{firstpage}--\pageref{lastpage}} \pubyear{2010}

\maketitle

\label{firstpage}

\begin{abstract}
We investigate the effects of potential sources of systematic error on the angular and photometric redshift, $z_{phot}$, distributions of a sample of redshift $0.4 < z < 0.7$ massive galaxies whose selection matches that of the Baryon Oscillation Spectroscopic Survey (BOSS) constant mass sample. Utilizing over 112,778 BOSS spectra as a training sample, we produce a photometric redshift catalog for the galaxies in the SDSS DR8 imaging area that, after masking, covers nearly one quarter of the sky (9,913 square degrees). We investigate fluctuations in the number density of objects in this sample as a function of Galactic extinction, seeing, stellar density, sky background, airmass, photometric offset, and North/South Galactic hemisphere. We find that the presence of stars of comparable magnitudes to our galaxies (which are not traditionally masked) effectively remove area. Failing to correct for such stars can produce systematic errors on the measured angular auto-correlation function, $w(\theta)$, that are larger than its statistical uncertainty. We describe how one can effectively mask for the presence of the stars, without removing any galaxies from the sample, and minimize the systematic error. Additionally, we apply two separate methods that can be used to correct the systematic errors imparted by any parameter that can be turned into a map on the sky. We find that failing to properly account for varying sky background introduces a systematic error on $w(\theta)$. We measure $w(\theta)$, in four $z_{phot}$ slices of width 0.05 between $0.45 < z_{phot} < 0.65$ and find that the measurements, after correcting for the systematic effects of stars and sky background, are generally consistent with a generic $\Lambda$CDM model, at scales up to 60$^{\rm o}$. At scales greater than $3^{\rm o}$ and $z_{phot} > 0.5$, the magnitude of the corrections we apply are greater than the statistical uncertainty in $w(\theta)$. The photometric redshift catalog we produce will be made publicly available at http://portal.nersc.gov/project/boss/galaxy/photoz/.
\end{abstract}

\begin{keywords}
Galaxies -- clustering.
\end{keywords}

\section{Introduction}
Wide-field, multi-band imaging surveys provide photometric redshift estimates for many millions of galaxies. Photometric redshifts, $z_{phot}$, are easier to obtain than spectroscopic ones, $z_{spec}$, but the gain in numbers of objects is countered by redshift uncertainties, $\sigma_z$, that are rarely better than $\sigma_z = 0.03(1+z)$. Such photometric redshift surveys may be referred to as having ``2 + 1'' dimensions --- nearly all of the radial clustering information is lost, but the redshift errors are small enough to allow two dimensional clustering measurements in redshift shells of width similar to $\sigma_z$. This strategy has been utilized to explore the formation and evolution of galaxies (see, e.g., \citealt{blakehod,M08,R09,R10,R10red,wakenewfirm}) and quasars (e.g., \citealt{Myers06}) and also to measure cosmological parameters (see, e.g., \citealt{blakecl,Pad07,R08,thomasneut,thomas10,CrocceDR7}). Such studies are gaining in importance, as future surveys such as The Dark Energy Survey\footnote{http://www.darkenergysurvey.org} (DES), The Large Synoptic Survey Telescope\footnote{http://www.lsst.org/lsst} (LSST), and the Panoramic Survey Telescope \& Rapid Response System\footnote{http://pan-starrs.ifa.hawaii.edu/public} (Pan-STARRS) will rely primarily on photometric redshifts.

At the largest scales, the accuracy of clustering measurements is not critically dependent on the uncertainty of each $z_{phot}$. However, contaminants or incorrect calibrations do matter at these scales --- the predicted clustering amplitudes are negligible; systematic errors can cause small fluctuations and thus non-zero amplitudes. Studies (see, e.g., \citealt{saw09,thomassys}) have found apparent excesses in the clustering strength at scales larger than 100 $h^{-1}$Mpc. Thorough studies are thus necessary to determine any potential sources of systematic error that could cause spurious fluctuations in the galaxy density field. 

In this paper, we investigate the observational realities that may cause fluctuations in the observed density of galaxies when modelled incorrectly. These include stellar contamination and masking, Galactic extinction, sky brightness, seeing, airmass, and offsets in photometric calibration. The effect of stellar contamination in a galaxy sample is well documented (see, e.g. \citealt{Myers06,thomassys,CrocceDR7}). Stars may also cause a systematic effect on the number density of objects by occulting a small fraction of the sky. This area is on the order of 1 millionth of a square degree per star, but with tens of millions of stars, becomes substantial given the precision to which clustering measurements can now be made. 

Galactic extinction requires that magnitudes be corrected for the effect of dust in our Galaxy. It has been noted several times (e.g., \citealt{Scranton02,Myers06,R06,Ho08}; Wang \& Brunner in prep.) that errors in this correction may cause a systematic effect on the galaxy density field, as the effective depth of a survey would fluctuate. Further, constant (extinction corrected) magnitudes have different fluxes (since the flux is directly related to the magnitude {\it before} extinction corrections).  This implies that the expected magnitude error will vary as a function of the Galactic extinction. Airmass has a similar
effect --- this simply refers to the path length of the photons through
our atmosphere to the telescope, normalized to unity for observations
at the zenith where it is minimized. At higher airmass less photons
reach the detector because more are scattered/absorbed in the
atmosphere and thus the error on a measured magnitude will be related to the airmass. Finally, the observed flux of an object is more spread out at higher seeing --- this increases the magnitude error and makes it more difficult to distinguish between stars and galaxies. Either of these seeing-dependent effects could cause spurious fluctuations in the observed density of galaxies.

We use data from the Sloan Digital Sky Survey (SDSS; \citealt{York}) eighth data release (DR8; \citealt{DR8}) to identify and remove potential sources of systematic error on the angular clustering of objects selected to be luminous galaxies (LGs) with redshifts $0.4 < z < 0.7$. Section \ref{sec:data} presents the data we use for our photometric redshift catalog and the spectroscopic data we use to train the photometric redshifts we generate. Section \ref{sec:mmc} describes how we measure and model angular correlation functions. In Section \ref{sec:tar} we investigate the fluctuations we find in the observed number density of LGs as a function of observational parameters and correct for the systematic errors these variations may impart. Section \ref{sec:spec} explains how we train the photometric redshifts, the potential systematic effects we consider for this training, and the resulting photometric redshift catalog that we generate. In Section \ref{sec:wmeas}, we present  measurements of angular auto- and cross-correlation functions in slices of width $\Delta z_{phot} = 0.05$, test their consistency with a generic $\Lambda$CDM model, and determine how the galaxy bias we calculate changes depending on the corrections we apply. We conclude with a summary of our results and a discussion of its greater implications in Section \ref{sec:con}.

\section{Data}
\label{sec:data}
\begin{figure*}
\begin{minipage}{7in}
\centering
\includegraphics[width=180mm]{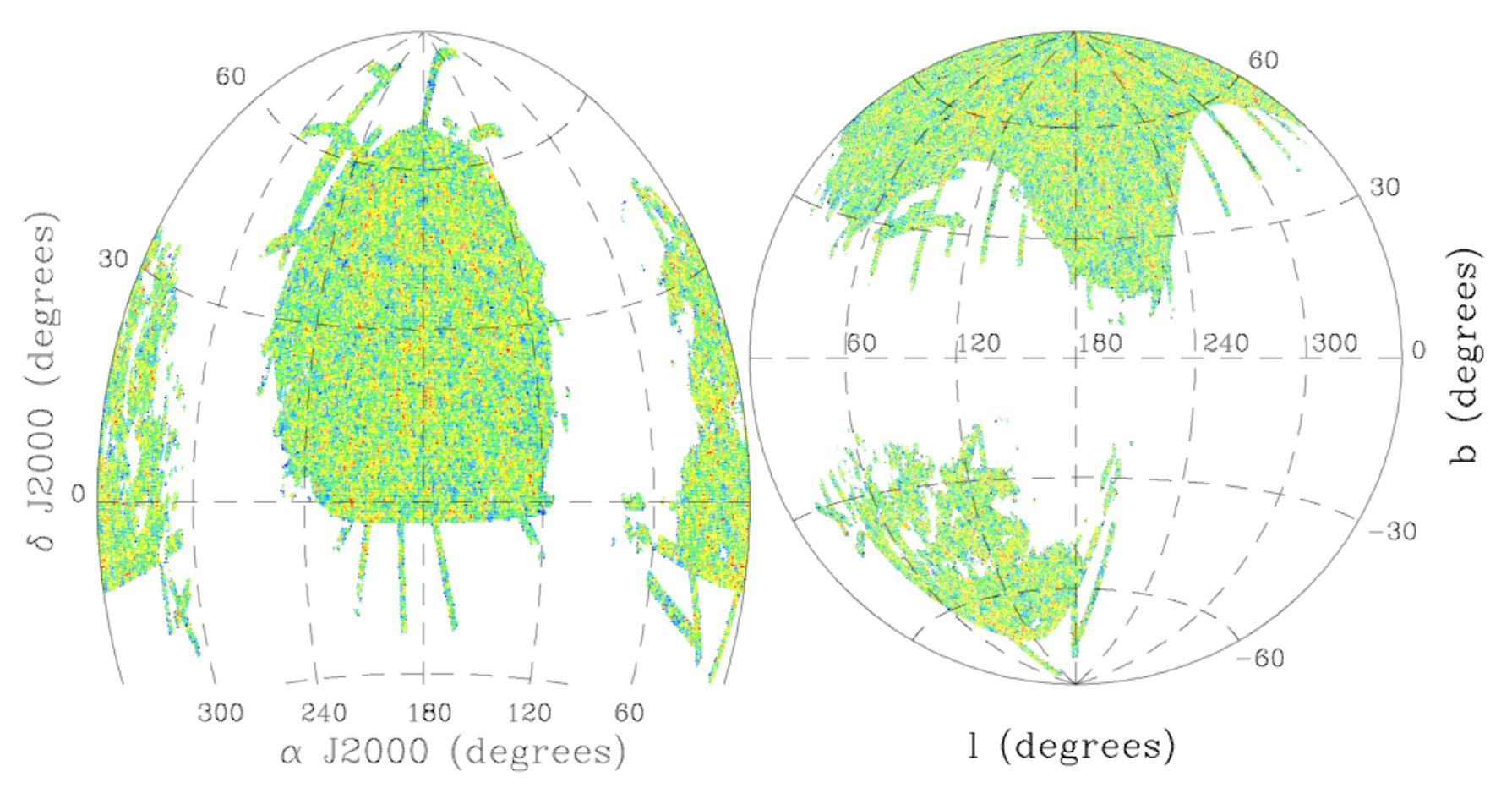}
\caption{The density of objects in our catalog, in equatorial coordinates (left panel) and Galactic coordinates (right panel), after masking for imaging area, seeing, Galactic extinction, and bright stars. This masked footprint occupies 9,913 square degrees. The density increases from blue to red, with blue representing a density that is less than 40\% of the average and red representing a density that is 120\% greater than average. }
\label{fig:footprint}
\end{minipage}
\end{figure*}

We use imaging data from the SDSS DR8 (\citealt{DR8}) to create a photometric redshift catalog of galaxies. This survey obtained wide-field CCD photometry (\citealt{C,Gunn06}) in five passbands ($u,g,r,i,z$; e.g., \citealt{F}), amassing a total footprint of 14,555 deg$^2$ data for which object detection is reliable to $r \sim 22$ (\citealt{DR8}). 

We use spectroscopy from the SDSS III Baryon Oscillation Spectroscopic Survey (BOSS;\citealt{Eisenstein11}) to obtain the spectroscopic redshifts, $z_{spec}$, we use as a training sample for our photometric redshift catalog. BOSS is a spectroscopic survey that will target 1.5 million massive galaxies, 150,000 quasars, and over 75,000 ancillary targets over an area of 10,000 sq. degrees \citep{Eisenstein11}. BOSS observations began in 2009, and the last data will be acquired in 2014. The BOSS spectrographs (R = 1300-3000) are fed by 1000 fibres in a single pointing, each with a 2$^{\prime\prime}$ aperture. Each observation is performed in a series of 15-minutes exposures and integrated until a fiducial minimum signal-to-noise is reached. This insures an isotropic sample, complete to high redshift ($z\sim0.7$), resulting in a redshift completeness of $\sim$ 97\% over the full imaging footprint. 

\subsection{Selecting Imaging and Redshift Data}
Our photometric catalog has the same selection as the sample of BOSS targets chosen to have approximately constant stellar mass, denoted `CMASS', as described by \cite{Eisenstein11}. We select objects from the Catalog Archive Server (CAS) PhotoPrimary table\footnote{see http://skyserver.sdss.org/dr8/en/help/browser/browser.asp for descriptions of the data contained within this table} identified as galaxies. We designate the subscript $_{mod}$ to denote the SDSS uber-calibrated model magnitudes \citep{Pad08}. The subscript $_{cmod}$ denotes cmodel magnitudes, where the cmodel flux, $\Phi_{cmod}$, is defined as:
\begin{equation}
\Phi_{cmod} = f_{psf}\Phi_{dev} + \Phi_{exp}(1.0-f_{psf})
\end{equation}
where $\Phi$ is the flux, subscripts $_{dev}$ and $_{exp}$ refer to the best-fit DeVaucouleurs and exponential profiles, and $f_{psf}$ is fraction of the flux within the PSF. The CMASS selection is then defined by:
\begin{eqnarray}
%\begin{array}{l}
 17.5 < i_{cmod}  < 19.9\\
r_{mod} - i_{mod}  < 2 \\
d_{\perp} > 0.55 \\
i_{fiber2} < 21.7\\
i_{cmod}  < 19.86 + 1.6(d_{\perp} - 0.8) \label{eq:slide}
%\end{array}
\end{eqnarray}
where all magnitudes are corrected for Galactic extinction, $i_{fiber2}$ is $i$-band magnitude within a $2^{\prime \prime}$ aperture\footnote{We use an $i_{fiber2}$ limit of 21.7; although the limit has changed to 21.5 in current BOSS targeting, the limit was $\leqslant 21.7$ for all of the BOSS spectra in this study}, and 
\begin{equation}
d_{\perp} = r_{mod} - i_{mod} - (g_{mod} - r_{mod})/8.0 .
\label{eq:dp}
\end{equation}

Stars are further separated from galaxies by only keeping objects with
\begin{eqnarray}
%\begin{array}{l}
i_{psf} - i_{mod} > 0.2 + 0.2(20.0-i_{mod})\\
z_{psf}-z_{mod} > 9.125 -0.46z_{mod}
\end{eqnarray}
\label{eq:sgsep}
%\end{equation}
unless the object passes a `LOWZ' cut defined by
\begin{eqnarray}
%\begin{array}
r_{cmod} < 13.6 + c_{\parallel}/0.3\\
|c_{\perp}| < 0.2 \\
16 < r_{cmod} < 19.6 
\end{eqnarray}
%\end{equation}
where
\begin{equation}
c_{\parallel} = 0.7(g_{mod}-r_{mod})+1.2(r_{mod}-i_{mod}-0.18)
\end{equation}
and
\begin{equation}
c_{\perp} = r_{mod}-i_{mod}-(g_{mod}-r_{mod})/4.0 -0.18 .
\end{equation}

These target selection criteria produce a sample of just over 1.6 million objects, occupying over 11,000 square degrees of area on the sky. We refer to these objects as `LG' (for luminous galaxy). We cut this sample down to the main SDSS imaging area. We define this area as the data contained in {\rm HEALPix} \citep{Gor05} pixels at N$_{\rm side}$ = 1024 (this resolution breaks the full-sky into 12,582,912 equal area pixels). Each pixel is assigned a weight given its overlap with the imaging footprint (accounting for the area taken up by bright stars), and we include only pixels with weight at least 0.9. This process is described in detail in Ho et al. (2011). Further, we only use data with seeing (defined by the $r$-band psf-{\rm FWHM}) less than 2$^{''}$.0 and Galactic extinction, $E(B-V) < 0.08$, as determined via the dust maps of \cite{SFD}. These cuts remove a large fraction of the data, leaving only 1,065,823 objects.  Their footprint is displayed in Fig. \ref{fig:footprint}: 9,913 square degrees, 2,554 of which are in the southern stripes. A total of 282,687 of the objects are in the southern stripes, meaning their number density is 110.7 deg$^{-2}$, while the number density in the north is 107.1 deg$^{-2}$ --- a 3.4\% difference. 

We match the masked LGs to the BOSS CMASS spectra that had been observed and run through the spectroscopic pipeline up to the 11$^{th}$ of November, 2010. This yielded a sample of 112,778 spectroscopic redshifts that are used to estimate photometric redshifts for our full sample. We find that 3.7\% of these spectroscopic objects are either stars or quasars (2.7\% stars and 1\% quasars). The percentage of quasars should be roughly constant across the sky, but the stellar contamination will be highly dependent on the proximity to the Galactic disk and center. We find that the percentage of stars varies from 6\% at Galactic latitude $b = 25$ to 1\% at $b = 85$. Masters et al. (2011) find 2+/-1\% point source contamination by inspecting high resolution Hubble Space Telescope images of BOSS CMASS
targets in the COSMOS survey field (at $b=42^{\rm o}$).

We also use stars to investigate systematic effects. We select objects from PhotoPrimary that are identified as stars and have $17.5 < i_{mod} < 20.5$. In total, this is over 84 million objects, but only 33 million reside within our masked footprint.

\subsection{Star/Galaxy Probability}
\label{sec:sgsep}
For the sample of BOSS spectra we use, 3.7\% of objects that are targeted as CMASS galaxies are spectroscopically classified as either stars or quasars. We use the software package ANNz\footnote{http://www.homepages.ucl.ac.uk/$\sim$ucapola/annz.html} \citep{F03} to identify stars, as in previous studies (e.g., \citealt{megaz}). Assigning galaxies a value of 1 and stars/quasars a value of zero, we divide our spectroscopic sample such that one quarter are placed into a training set, another one quarter into a validation set, and the remaining half into a testing set. We then train ANNz in order to classify galaxies, using the five SDSS model magnitudes and the parameters $i_{cmod}$, $i_{psf}$,$ i_{fiber2}$, $i_{exp}$, $R_{pet,i}$, $R_{dev,i}$, $R_{exp,i}$, $AB_{dev,i}$, $AB_{exp,i}$, ln$L_{star}$, ln$L_{exp}$, ln$L_{dev}$, where $R$ is the radius, $AB$ is the axis ratio, subscript $_{pet}$ refers to the best-fit Petrosian profile, and `ln$L$' stands the natural log of the likelihood. The values ANNz returns, which we denote `$p_{sg}$', are predominantly between 0 and 1, and when they are outside of these bounds, we set them to 0 and 1, respectively. (This affects only 1\% of the objects and does not bias the over-all probability distribution.) 

\begin{figure}
\includegraphics[width=84mm]{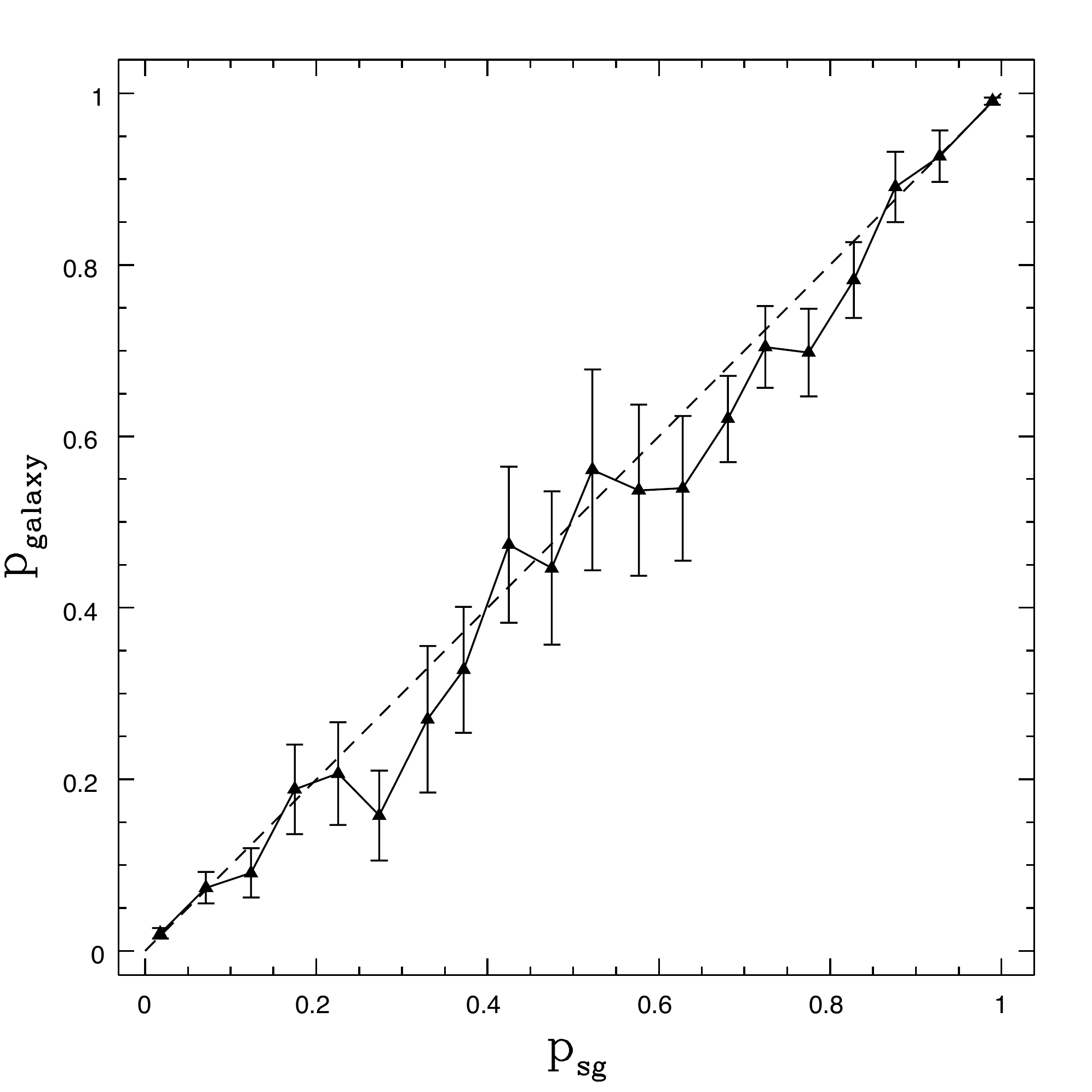}
\caption{The fraction of objects that are galaxies versus the value of the star/galaxy parameter ($p_{sg}$).  The dashed line displays the relationship $p_{galaxy} = p_{sg}$. Errors are Poisson.}
\label{fig:pgal}
\end{figure}

We find that the star/galaxy training also does a good job of estimating the probability that an object is a galaxy.  Fig. \ref{fig:pgal} displays the fraction of objects that are galaxies (`$p_{gal}$', as determined from their spectra) versus the value of the star/galaxy parameter, $p_{sg}$. These two quantities are nearly identical, as can be seen by comparing to the dashed line. This implies we can treat $p_{sg}$ as the probability that an object is a galaxy, allowing us to remove most of the effect of the stellar contamination. We note that the $p_{sg}$ estimation benefits greatly from having a large training set distributed over $21.3^{\rm o} < |b| < 83.5^{\rm o}$ (less than 4\% of the objects in our catalog lie outside these bounds). Unless otherwise noted, throughout we will be counting LGs by summing their values of $p_{sg}$. For our full (masked) sample, the sum of $p_{sg}$ is 1,021,885, suggesting that 4.1\% of the objects are stars or quasars. 

\section{Measuring and Modeling Correlation Functions}
\label{sec:mmc}
The primary focus of this work is to determine how systematics may affect the measured clustering signal. Therefore, we measure angular auto- and cross-correlation functions, $w(\theta)$, of the density fields of LGs and of potential systematics. These statistics can be calculated extremely quickly\footnote{Approximately 15 minutes total processing time using a single 2.53 GHz processor} and can be compared to cosmological models that are well tested by simulation.

\subsection{Estimating $w(\theta)$ and its Covariance}
\label{sec:cc}
We calculate $w(\theta)$ using {\rm Healpix} pixels with N$_{\rm side}$ = 256. Our mask has removed all data in pixels with weight $< 0.9$ at the N$_{\rm side}$ = 1024 resolution. Thus, given that the weights themselves are a good approximation of the fractional area of a pixel, the area per pixel for N$_{\rm side}$ = 256 is well approximated by its weight, $wt$, multiplied by the area of the pixel. Therefore the over-density in pixel $i$, $\delta_i$, is given by
\begin{equation}
\delta_i = n_{i}/(\bar{n}wt_i) - 1,
\end{equation}
where $n_i$ is the number of galaxies in pixel $i$. For LGs and stars, $\bar{n} = \sum n_i / \sum wt_i$, while for observational parameters (such as Galactic extinction), $\bar{n} = \sum A_iwt_i / \sum wt_i$, where $A_i$ would be the average value of the observational parameter in pixel $i$. The correlation function, $w(\theta)$, is given by (see, e.g., \citealt{Scranton02})
\begin{equation}
w_{a,b}(\theta) = \frac{\sum_{i,j}\delta^a_i\delta^b_j\Theta_{i,j}(\theta)wt_iwt_j}{\sum_{i,j}\Theta_{i,j}(\theta)wt_iwt_j},
\label{eq:wcalc}
\end{equation}
where $\Theta_{i,j}$ is equal to 1 if pixel $i$ is at an angular distance $\theta \pm \Delta\theta$ from pixel $j$ and zero otherwise, $a = b$ represents an auto-correlation function, and $a \ne b$ represents a cross-correlation function. 

In order to compare $w(\theta)$ measurements, we calculate jack-knife errors, $\sigma_{\rm jack}$. We use 20 equal area jack-knife regions. This is accomplished by selecting contiguous regions of Healpix pixels whose weights sum to 1/20th of the total. The jack-knife errors are defined by (see, e.g., \citealt{Scranton02,Myers07,R07})
\begin{equation}
\sigma^2_{\rm jack}(\theta) = \frac{19}{20}\sum_{i=1}^{i=20}\left[w(\theta)-w_i(\theta)\right]^2,
\end{equation}
where $w(\theta)$ is the measurement over the entire area and $w_i(\theta)$ is the measurement when the $i$th jack-knife region is removed. 

We calculate the covariance matrix that we use to compare our $w(\theta)$ measurements to theoretical models by transforming theoretical $P(k)$, to angular power-spectra, and from the covariance of the angular power spectra to the $w(\theta)$ covariance, $C(\theta_1,\theta_2)$. The methods for doing this are outlined in, e.g., \cite{crocce10} and \cite{RSDme}. Specifically, $C(\theta_1,\theta_2)$ is given by Eqs. 13 - 16 of \cite{RSDme}, using a linear $P(k)$. \cite{crocce10} have extensively tested these errors against mock catalogs, finding them to accurately reproduce the covariance in $w(\theta)$ at linear scales, and we restrict this study to those scales. Further, at the large scales, estimations of the off-diagonal elements of the covariance matrix constructed using jack-knife methods have large statistical uncertainty and can therefore lead to unstable covariance matrices. Thus, we compare our measured $w(\theta)$ to the model $w(\theta)$ and $C(\theta_1,\theta_2)$ in order to find the best-fit model. In Section \ref{sec:wmeas}, we compare the theoretical and jack-knife uncertainties.

\subsection{Modeling}
\label{sec:th}
We compare our measurements to theory, assuming a flat $\Lambda$CDM cosmology with $h = 0.7$, $\Omega_m = 0.274$, $f_b = \Omega_b/\Omega_m = 0.18$, $n =0.95$, and $\sigma_8 = 0.8$ (as used in \citealt{whiteboss}), and a CMB temperature of 2.725 K. We calculate the $z=0$ linear power-spectrum using CAMB\footnote{see camb.info} \citep{camb}. We account for the effects of structure growth via (see, e.g., \citealt{Seo05,CS06}):
\begin{equation}
P(k,z) = D(z)^2P(k,0)e^{-\left[ks_bD(z)\right]^2}
\end{equation}
where $D(z)$ is the linear growth rate; the exponential term accounts for the damping imparted by large-scale velocity flows. We derive $s_b= 5.27h^{-1}$ Mpc using the Zel'dovich approximation
\citep{ESW07} for our fiducial cosmology. Such an approximation has been shown to be a good fit to the BAO feature in recent N-body simulation results (see, e.g., \citealt{Seo10}).
\footnote{This damping scale corresponds to 7.45$h^{-1}$ Mpc, using the
convention used in \cite{ESW07}
.}

Given $P(k,z)$, we Fourier transform to determine the isotropic 3-dimensional real-space correlation function $\xi_{lin}(r)$, (which is implicitly dependent on redshift). We then incorporate the effects of mode-coupling (see, e.g., \citealt{crocce08,Sanchez08,crocce10}) via
\begin{equation}
\xi(r) = \xi_{lin}(r)+A_{mc}\xi^{(1)}_{lin}(r)\xi^{\prime}_{lin}(r),
\end{equation}
where $\xi^{\prime}_{lin}$ is the derivative of $\xi_{lin}$,
\begin{equation}
\xi_{lin}^{(1)} \equiv \frac{1}{2\pi^2} \int P(k,z)j_1(kr)kdk,
\end{equation}
and for $A_{mc}$ we use the value of 1.55 determined by \cite{crocce10}.

We model the redshift-space correlation function as \citep{hamilton92}
\begin{equation}
\xi^s(\mu,r) = \xi_o(r)P_o(\mu)+\xi_2(r)P_2(\mu) +  \xi_4(r)P_4(\mu),
\label{eq:ximus}
\end{equation}
where
\begin{eqnarray}
  \xi_0(r) &=& (b^2+\frac{2}{3}bf+\frac{1}{5}f^2)\xi(r), \\
  \xi_2(r) &=& (\frac{4}{3}bf+\frac{4}{7}f^2)[\xi(r)-\xi'(r)], \\
  \xi_4(r) &=& \frac{8}{35}f^2[\xi(r)+\frac{5}{2}\xi'(r)-\frac{7}{2}\xi''(r)],
\label{eq:xirs}
\end{eqnarray}
$P_\ell$ are the Legendre polynomials,
\begin{eqnarray}
  \xi'\equiv3r^{-3}\int^r_0\xi(r')(r')^2dr' \\ 
  \xi''\equiv5r^{-5}\int^r_0\xi(r')(r')^4dr',
\end{eqnarray}
$b$ is the large-scale bias of the galaxy population being considered, and $\mu$ is the cosine of the angle to the line of sight. In order to calculate model $w(\theta)$, we must project $\xi^s(\mu,r)$ over the radial distribution of galaxy pairs in a particular sample (or samples in the case of cross-correlations).
\begin{equation} 
 \label{eq:w2}
  w(\theta) = \int dz_1 \int dz_2 n_{i}(z_1)n_{j}(z_2)
    \xi^s\left[\mu,{\bf r}(\theta,z_1,z_2)\right],
  \label{eq:wnz}  
\end{equation}
where $n_{i}$ is the normalized redshift distribution of sample $i$ (and $i = j$ for the auto-correlation).  The galaxy separation ${\bf r}$ is a function of the angular separation of the galaxies $\theta$ and their redshifts $z_1$ and $z_2$ (as is $\mu$).

\subsection{Correcting Spurious Clustering}
\label{sec:wcorr}
Observational effects may cause spurious fluctuations in the number of observed LGs. As first derived (for large scale clustering) in Ho et al. (in prep), to first order, the systematic effect contributed by $i$ observational parameters on the observed density field is given by 
\begin{equation}
\delta^{\rm o}_g = \delta^{t}_g+\sum_i\epsilon_i\delta_i.
\label{eq:dgds}
\end{equation}
\noindent where $\delta^{\rm o}_g$ is the over-density of galaxies we observe, $\delta_i$ is the over-density of the systematic $i$, $\delta^{t}_g$ is the true over-density of galaxies, and $\epsilon_i$ assumes a linear relationship between the potential systematic and its effect on the observed over-density of galaxies. According to Eq. \ref{eq:wcalc}, $w(\theta) = \langle \delta_i\delta_j\Theta_{i,j}(\theta)\rangle$. Thus,
\begin{equation}
w^{t}_g(\theta) =w^{\rm o}_g(\theta) - \sum_i \epsilon_i^2 w_i(\theta)- \sum_{i,j>i}2\epsilon_i\epsilon_j w_{i,j}(\theta)
\label{eq:acc}
\end{equation}
and 
\begin{equation}
w^{o}_{g,i} =  \sum_{j}\epsilon_j w_{i,j}(\theta) ,
\label{eq:ccc}
\end{equation}

\noindent We can measure $w^{o}_g(\theta)$ (the auto-correlation function of our galaxy sample), $w^{o}_{g,i}$ (the cross-correlation function of our galaxy sample with systematic $i$), and $w_{i,j}(\theta)$ (the auto-/ cross-correlation function of the systematics). Thus we always have $i+1$ equations and unknowns ($\epsilon_i$ and $w^{t}_g(\theta)$) and we can therefore solve for $w^{t}_g(\theta)$. We present the solutions for three systematics in Appendix \ref{app:sysc}. We note that measuring the cross-correlation between observational parameters and galaxies in order to identify potential systematic errors has been applied to SDSS data since its early-data release \citep{Scranton02}.

\section{Systematic Effects on the Angular Distribution of Galaxies}
\label{sec:tar}
\begin{figure*}
\begin{minipage}{7in}
\centering
\includegraphics[width=180mm]{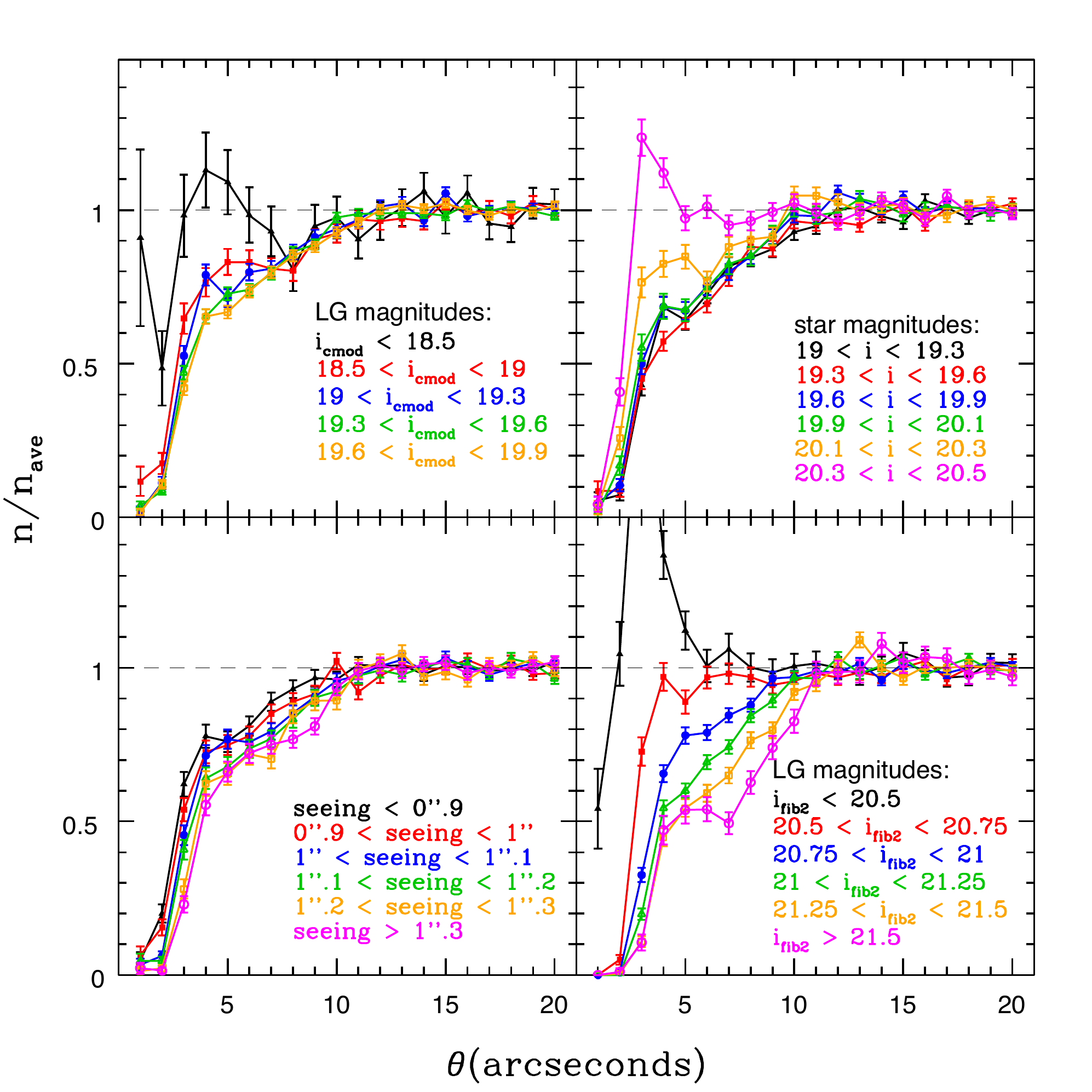}
\caption{The number density of galaxies we select around stars with $17.5 < i_{mod} < 19.9$ (unless otherwise noted), divided by the average number density, in 1$^{\prime\prime}$ wide annuli, plotted against the maximum radius of the annulus. The top-left panel displays the results when we divide the galaxy sample into the noted $i_{cmod}$ magnitude bins. The top-right panel displays the case where we use the full galaxy sample and find the number densities around stars within the noted $i$-band magnitude bins. In the bottom left panel, we display the result when restrict to imaging regions with the labeled seeing. In the bottom-right panel, we divide the LGs into bins based on their $i$-band magnitude within a 2$^{\prime\prime}$ aperture ($i_{fib2}$). Errors are Poisson.}
\label{fig:ntararoundstar}
\end{minipage}
\end{figure*}

Effects such as seeing, Galactic extinction, and sky brightness may affect the number density of galaxies we select. Another important consideration is the presence of foreground objects. As shown in Fig. 4 of \cite{DR8}, the presence of foreground objects has a significant effect on the number density of background objects one is able to detect. 

\subsection{Foreground Stars}
\label{sec:fst}

In order to investigate the effect of foreground stars on the observed density of LGs, we determine the number density of LGs in the immediate vicinity of stars within our masked footprint. In annuli of width 1$^{\prime\prime}$ around each star, we determine the number density of LGs as a function of the maximum radius of the annulus. In the top left panel Fig. \ref{fig:ntararoundstar}, we present this measurement when considering stars with $17.5 < i_{mod} < 19.9$ and dividing the LG sample into five $i_{cmod}$ bins. We find that the presence of a star has a significant effect on the ability to observe LGs with $i_{cmod} > 18.5$ out to at least 10$^{\prime\prime}$ (we note that there are only 22,000 LGs with $i_{cmod} < 18.5$, making the results in this bin relatively uncertain). The effect remains nearly constant as a function of the $i_{cmod}$ magnitude, but it is strongest for the faintest sample (displayed in orange).

We also find that for $i < 20.1$, the $i$-band magnitude of the star does not produce a strong effect. This is shown in the top-right panel of Fig. \ref{fig:ntararoundstar}, where we take the full LG sample and find the number density of these objects around six separate samples of stars that we have divided, based on model magnitudes, into bins between $19 < i < 20.5$. We find similar results for the bins with $19 < i < 20.1$ (the black, red, blue, and green points and lines). The effect becomes significantly weaker for stars with $20.1 < i < 20.3$. For the $20.3 < i < 20.5$ bin, the effect is removed when the annuli have outer radii at least 5$^{\prime\prime}$. When the outer radii are 3$^{\prime\prime}$ and 4$^{\prime\prime}$, there is an excess of LGs. It is possible that this excess is caused by (resolved) binary stars --- many stars are members of binary systems and thus there is an enhanced likelihood that an object within a few arcseconds of a star is also a star, and therefore the stellar contamination rate in our LG sample will be higher around stars. 

We find a deficit of LGs around stars with $i_{mod} < 20.3$ to at least 10$^{\prime\prime}$ from the star. This suggests that the extended seeing disc of a star increases the sky noise in its vicinity and therefore makes object detection less likely. This implies that the effect should depend both on the seeing during the observation and the surface brightness of the object that might be detected. In the bottom-left panel of Fig. \ref{fig:ntararoundstar}, we use the full LG sample and stars with $17.5 < i < 19.9$ and divide the imaging area into six regions based on seeing (we note the median seeing in our masked footprint is 1$^{\prime\prime}$.07). As expected, we find that the deficit of LGs close to stars becomes greater as the seeing becomes more poor. However, there is still a significant deficit of LGs close to stars at all levels of seeing.

In the bottom-right panel of Fig. \ref{fig:ntararoundstar}, we find the number density of LGs around stars with $17.5 < i < 19.9$ when we divide the LGs into six bins based on the $i_{fib2}$ magnitude of the LG. We find dramatic differences, as we find a large {\it excess} for the brightest LGs (black, $i_{fib2} < 20.5$) and the largest deficit for the faintest LGs (magenta, $i_{fib2} > 21.5$). The $i_{fib2}$ magnitudes are a measure of the flux within a constant aperture, and are therefore a measure of the surface brightness of the LG. Thus, as expected, we find that the presence of a star has the greatest effect on the objects with the lowest surface brightness. We find a large excess of LGs with $i_{fib2} < 20.5$ between 2$^{\prime\prime}$ and 5$^{\prime\prime}$ from stars. We believe this excess caused by binary companions to the stars we are testing against, as the most compact objects will have the highest surface brightness (at constant $i_{mod}$) and are also most likely to be morphologically similar to stars.

\begin{figure*}
\begin{minipage}{7in}
\includegraphics[width=7in]{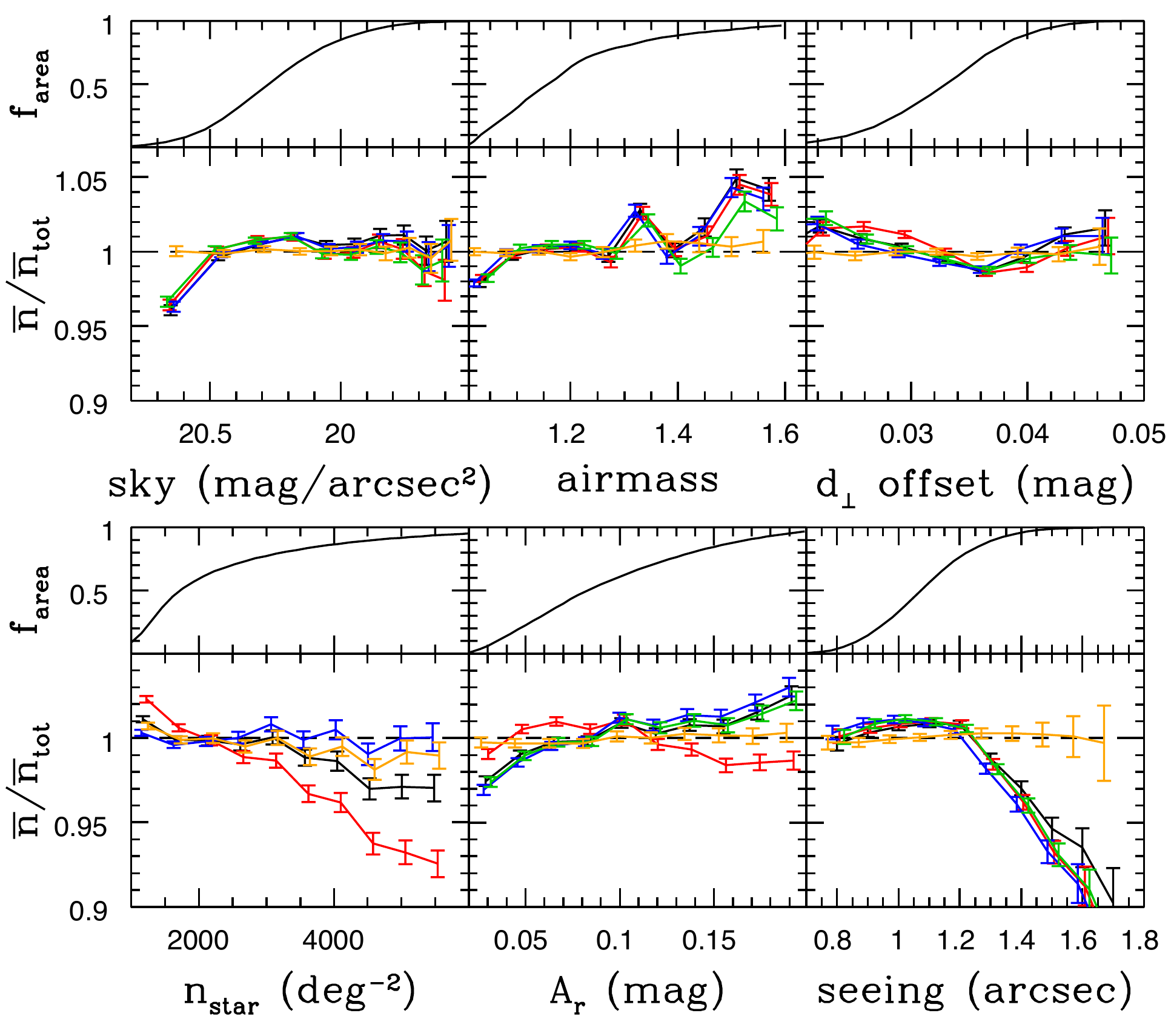}
\caption{Each of the six panels is divided such that the bottom portion displays the number density, divided by the average number density, as a function of the observational parameter. Individual points are connected by solid lines and represented by error-bars, whose size are calculated assuming Poissonian statistics. Black represents equally weighted LGs (which can be both galaxies and stars) and red weighting each LG by the probability it is a galaxy, $p_{sg}$. The results plotted in blue show the scenario when we weight each LG by $p_{sg}$ and subtract the effective area of stars, $A_{star}$, from each pixel for each star. Green displays the case where we change the selection criteria to $d_{\perp} > 0.5564$ for objects in the South, weight each LG by $p_{sg}$, and subtract $A_{star}$. Orange represents the application of iterative weights to the LG density field, in an attempt to remove all of the fluctuations. The top portions of each panel display the fraction of the imaging area where the observational parameter is less than the value on the x-axis. From top-left to bottom right, the observational parameters are: sky background in $i$-band magnitudes/arcsec$^2$ (sky); $i$-band airmass (airmass); the estimated offset in $d_{\perp}$ in magnitudes ($d_{\perp}$ offset); number density of stars with $17.5 < i < 19.9$ ($n_{star}$); $r$-band Galactic extinction in magnitudes ($A_r$); and $i$-band psf fwhm in arcseconds (seeing).  Errors are Poisson.}
\label{fig:ntarall}
\end{minipage}
\end{figure*}

Fig. 3 suggests that every star effectively removes a small amount of imaging area. If this effect is not corrected for, we would expect an anti-correlation between LGs and stellar density. However, 4\% of the objects in our catalog are stars, implying that, with no correction, there should be a positive correlation between LG and stellar density. The bottom left panel of Fig. \ref{fig:ntarall} presents the relationship between LG density and the density of stars selected from DR8 with $17.5 < i_{mod} < 19.9$. When we equally weight every object (black), we see a slight decrease ($\sim$3\% across the full range) in the number of objects as the stellar density increases. This suggests the foreground presence of stars (which removes objects from our catalog) dominates over the increase in objects we select (erroneously) as galaxies due to stellar contamination. If we instead weight each object by the probability that an object is a galaxy, $p_{sg}$ (as estimated in section \ref{sec:sgsep}; red), we find a significant and monotonic decrease (totalling 10\%) in the number of LGs as a function of stellar density. The upper portion of the bottom-left panel of Fig. \ref{fig:ntarall} displays the fraction of our (masked) imaging area where the number density of stars is below $n_{star}$. It shows that the majority of the data (60\%) has $n_{star} < 2000 ~{\rm deg}^2$, but there are still data (5\%) with $n_{star} > 6000~ {\rm deg}^2$. 

Foreground stars appear to remove area from the survey. Integrating $2\pi(1-n/n_{ave}(\theta)) \theta$, we can estimate the effective area lost per star due to the occultation effect. For the stars with $19.3 < i < 19.6$, this yields an effective area of 67.2 square arcseconds and thus an effective radius of 4$^{\prime\prime}$.6. 

Alternatively, we can assume that each star removes an effective area, which we denote `$A_{star}$', and we determine $A_{star}$ by finding the radius, $r_{star}$, which makes the values displayed in the bottom-left panel of Fig. \ref{fig:ntarall} closest to 1. We find that the $\chi^2$, using Poisson errors and the model $\bar{n}/\bar{n}_{tot} = 1$, is minimized for $r_{star} = 9^{\prime\prime}$.48. This $r_{star}$ is determined using only stars with $17.5 < i < 19.9$. Fig. 3 suggests that stars with $i$-band magnitudes as faint as 20.3 have an effect, and there are an additional 6.3 million stars with $19.9 < i < 20.3$ within our footprint. Scaling $r_{star}$ to account for these additional stars yields an effective circular area of radius of $8^{\prime\prime}.44$. This is still far greater than the value of $\sim5^{\prime\prime}$ we expect based on integrating $2\pi(1-n/n_{ave}(\theta)) \theta$ given the $n/n_{ave}(\theta)$ relationships displayed in Fig. \ref{fig:ntararoundstar}. Thus, the $\bar{n}/\bar{n}_{tot}(n_{star})$ relationship (displayed in the bottom-left panel of Fig. \ref{fig:ntarall}) is stronger than one might expect, suggesting there are additional effects due to stellar density beyond the occultation effect. This issue is studied in further detail in \cite{Pspec}.

We proceed by assuming each star effectively masks an amount of area consistent with $\bar{n}/\bar{n}_{tot}(n_{star}) = 1$. For our full LG sample and stars $17.5 < i < 19.9$, we determined $r_{star} = 9^{\prime\prime}$.48. This radius implies that stars are effectively removing a total area of 500 square degrees, which is 5\% of our masked footprint. The resulting $\bar{n}/\bar{n}_{tot}(n_{star})$ relationship is displayed in blue in the bottom-left panel of Fig. \ref{fig:ntarall}. We note that this effective radius is likely to depend on the magnitudes of the LGs, so any subsets of the data are likely to have different $r_{star}$. 

\begin{figure}
\includegraphics[width=84mm]{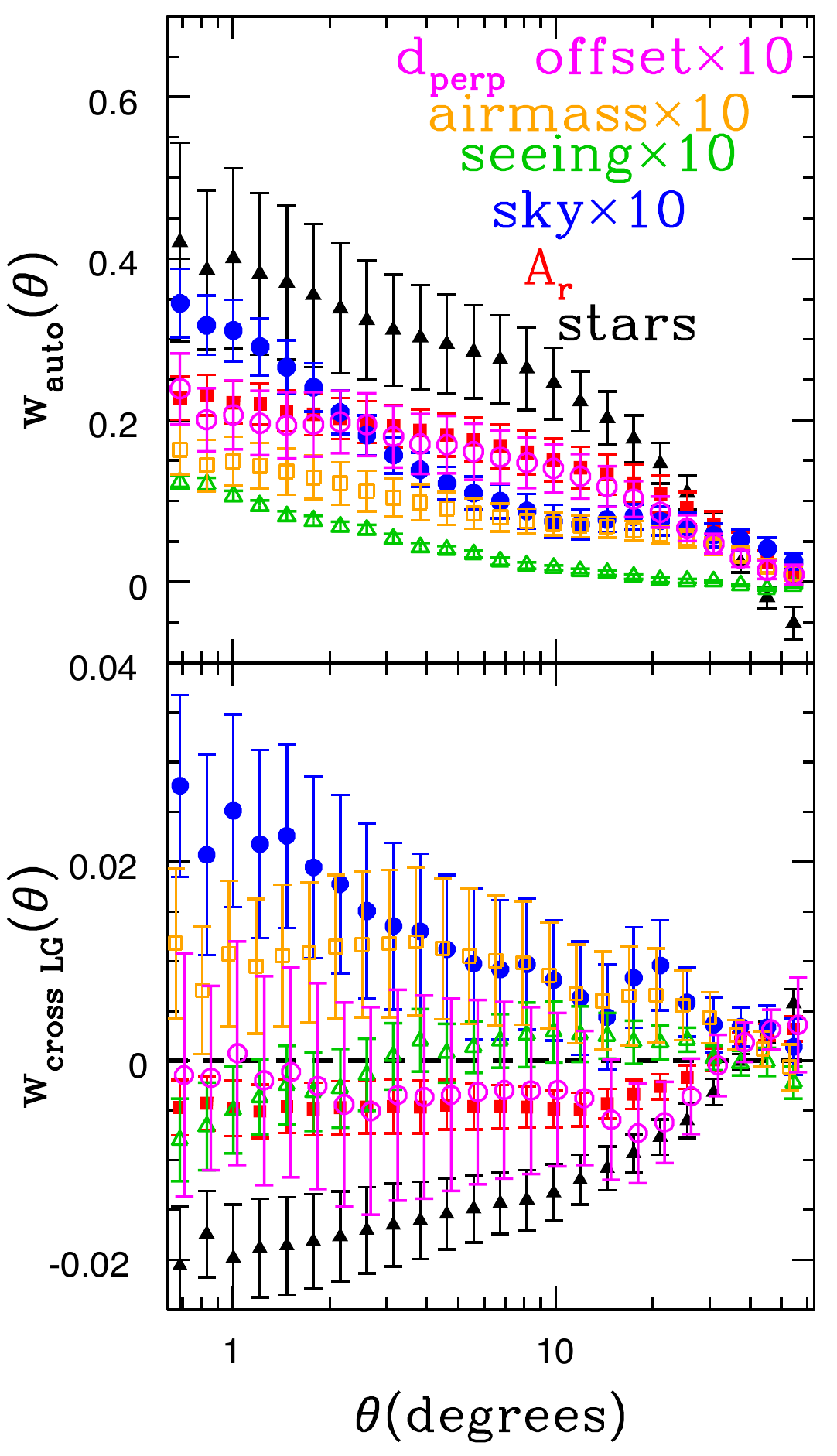}
\caption{Top panel: The auto-correlation functions of the density field of the stars (black triangles), $r$-band extinction ($A_r$; red squares), $i$-band sky background (sky; blue circles), $i$ band psf fwhm (seeing; green open triangles), $i$-band airmass (orange open squares), and the estimated offset in $d_{\perp}$ (magenta open circles). Bottom panel: The same observables, but cross-correlated with the $p_{sg}$ weighted LGs. In both panels, the error-bars are the estimated jack-knife errors, and the results using sky, seeing, airmass, and $d_{perp}$ have been multiplied by 10.}
\label{fig:wsys}
\end{figure}

The relationship between galaxy density and stellar density is important, due to the fact that stars display significant clustering on large angular scales (see, e.g., \citealt{Myers06}); the stars may therefore affect the measured clustering of galaxies at large physical scales. The auto-correlation function, $w(\theta)$, calculated as described in Section \ref{sec:cc}, of stars (with $17.5 < i < 19.9$) is displayed in black triangles in the top panel of Fig. \ref{fig:wsys}. The amplitudes are significant, and exhibit a monotonic decrease from $\sim$0.4 at $\theta = 1^{\rm o}$ to $\sim 0$ at $\theta = 50^{\rm o}$. The cross-correlation of the stars with the $p_{sg}$ weighted LGs, displayed in black triangles in the bottom panel of Fig. \ref{fig:wsys}, is significant and negative and increases towards 0 in a manner that mirrors the decrease in the star auto-correlation function. This implies that if it is unaccounted for, the presence of stars will cause systematic errors on the measured large-scale clustering of LGs.

We note that foreground stars will be a problem for any current or future large-scale-structure survey, and the problem will only become more significant as limiting magnitudes are pushed fainter and there are thus more foreground objects that may have a masking effect. Foreground galaxies will cause the same problem, but will have a much smaller effect on the measured clustering, since at large-scales the {\em angular} clustering amplitudes of foreground galaxies are significantly smaller than those of stars.

\subsection{Observational Parameters}
We find that the number density of LGs varies with observational parameters such as the Galactic extinction, the seeing, and the brightness of the background sky. In Fig. \ref{fig:ntarall}, we display the number-density, divided by the total average number-density, of LGs as a function of the value of the potential systematic, for the cases where we equally weight each object (black), we weight each object by its value of $p_{sg}$ (red), and we  weight each object by its value of $p_{sg}$ and subtract an area $A_{star}$ from each pixel for every star in the pixel (blue). In the top portion of each panel, we display the fraction of imaging data occupying area which has a value of the potential systematic that is lower than the value on the x-axis, i.e., the median value of the potential systematic is at $f_{area} = 0.5$. The median and inner/outer quartile values, within our masked footprint, of Galactic extinction, seeing, sky background, and airmass are determined using this information and displayed in Table \ref{tab:obpar}.

\begin{table}
\caption{The median and inner/outer quartile values of observational parameters in our masked footprint.} 
\begin{tabular}{lll}
\hline
\hline
parameter & median & inner, outer quartile\\
\hline
$A_r$ & 0.082 mag & 0.053, 0.126 mag\\
seeing &  1.$^{\prime\prime}$0.7 & 0.$^{\prime\prime}$96, 1.$^{\prime\prime}$19 \\
sky background & 20.27 mag & 20.43, 20.11 mag/as$^2$ \\
airmass & 1.16 & 1.08, 1.26  \\
\hline
\label{tab:obpar}
\end{tabular}
\end{table}

The bottom middle panel of Fig. \ref{fig:ntarall} plots the relationship between the number density of LGs and the Galactic extinction in the $r$-band, $A_r$ (we use the $A_r$ values from the CAS, which are based on the \citealt{SFD} dust maps and the relationship $A_r = 2.751E(B-V)$). With equal weighting (black; we note these data are nearly indistinguishable from those represented in blue and green), the number density of LGs increases slightly as a function of $A_r$. The $A_r$ values and stellar densities are highly correlated (since they both trace the structure of our Galaxy), but this does result in $\bar{n}/\bar{n}_{tot}(A_r)$ resembling the $\bar{n}/\bar{n}_{tot}(n_{star})$ relationship. Interestingly, the relationship flattens when we weight each object by $p_{sg}$, but reverts to its original form when we additionally subtract $A_{star}$. From the top portion of the panel, we can see that that the values of $A_r$ vary smoothly between $0.03 < A_r < 0.2$ and that its median value is $A_r = 0.08$. 

The auto-correlation function of $A_r$ is displayed in red squares in the top panel of Fig. \ref{fig:wsys}. The amplitudes are significantly non-zero and show a similar trend to the stars. Interestingly, the amplitudes of the Galactic extinction $w(\theta)$ are significantly smaller than those of the star $w(\theta)$, suggesting there is more structure to the distribution of stars in the Galaxy than to the dust in the Galaxy. The cross-correlation function of $A_r$ with the $p_{sg}$ weighted LGs (red squares in the bottom panel of Fig. \ref{fig:wsys}) is negative (except at the largest scales), but consistent with zero at a majority of scales. Interestingly the absolute values of this cross-correlation function are very similar in amplitude to those of the cross-correlation function we measure between $A_r$ and the LGs when we subtract the effective area of stars --- even though the $\bar{n}/\bar{n}_{tot}(A_r)$ relationship displays a significant change between the two treatments, we find no evidence for a systematic effect on the measured clustering

We also see significant changes in the number density as a function of the seeing. This result is shown in the bottom-right panel of Fig. \ref{fig:ntarall}. There is a 9\% decrease in the LG density between regions with seeing $1^{''}.2$ and those with $1^{''}.6$. We note that this is less than 25\% of the imaging area, as the upper quartile of seeing within our footprint is $1^{''}.19$. The reason for the decrease in LG number density in poor seeing is that the star/galaxy separation cut applied to BOSS targeting, given by Eq. \ref{eq:sgsep}, effectively changes. Increasing the seeing causes the PSF and model magnitudes to converge. The result is that in regions of poor seeing the two magnitudes are more similar - not because the object is too point-like, but rather because the PSF is too extended - and the cut is more likely to reject objects.

The green open triangles in the top panel of Fig. \ref{fig:wsys} present the measured $w(\theta)$ of the seeing. Its amplitudes are significantly smaller than either stars or $A_r$, though it is non-zero. This may be unexpected, as regions of similar seeing should follow the scanning pattern in the sky. However, in regions of sky that were imaged multiple times (roughly 50\% of the DR8 footprint), the imaging with better seeing is chosen. This works to alleviate any large-scale pattern in the seeing within our footprint and also reduces the median seeing to $1^{\prime\prime}$.07. The cross-correlation function of the seeing and LGs is displayed in the bottom panel of Fig. \ref{fig:wsys}. The amplitudes are consistent with zero but transition from being negative at smaller scales to positive at larger scales.

The sky background has a complex relationship with measured flux errors, and we may therefore expect the number density of observed LGs to depend on the sky background. The top-left panel of Fig. \ref{fig:ntarall} displays a 5\% increase in LG density between regions with a $i$-band sky background\footnote{The CAS gives sky background values, $f$, in terms of the flux unit of `nanomaggies/arcsec$^{2}$', which we convert to magnitudes/arcsec$^{2}$, $m$, via $m = 22.5 - 2.5{\rm log}(f)$ as implied by http://data.sdss3.org/datamodel/glossary.html\#nanomaggies} of 20.7 magnitudes/arcsec$^{2}$ (mag/as$^{2}$) to those with a background of 20.5 mag/as$^{2}$. This implies that the observed trend may be due to an increase in the average magnitude error scattering more objects into than out of our sample. However, 70\% of the footprint has a sky brightness between 20.5 and 20.0 mag/as$^{2}$ (as shown in the top right panel of Fig. \ref{fig:ntarall}) and the fluctuations are only $\sim$ 1\% in this range.

The auto-correlation function of the sky background is displayed in blue circles in the top panel of Fig. \ref{fig:wsys}. It is significantly positive, but is only $\sim$1/20th of that of the stars. The cross-correlation of the sky background with the LGs, displayed in the bottom panel of Fig. \ref{fig:wsys}, is significantly positive and $\sim$ 1/10th as large as the auto-correlation function of the sky background. This is the largest ratio between the auto- and cross-correlation functions of any of the potential systematics we measure. This suggests that the increase in LG number density between 20.7 and 20.5 mag/as$^{2}$ is related to a significantly positive cross-correlation function.

For an object of given brightness, the number of photons that make it to the CCD depends on the airmass. One may therefore also expect that the magnitude error will depend on the airmass. The top-middle panel of \ref{fig:ntarall} displays the relationship between the number density of LGs and the airmass. We do not find smooth variations --- rather we find a sharp increase in the number density where the airmass is approximately 1.35 and where it greater than 1.5 and also a decrease where it is less than 1.05. This suggests that the fluctuations are not tied to the physical effect of the value of the airmass, but rather these specific values of the airmass are correlated with other observational parameters. 

The auto-correlation function of the airmass and its cross-correlation function with LGs (orange open squares displayed in Fig. \ref{fig:wsys}) are quite similar to those of the sky background, especially at scales greater than 4$^{\rm o}$. Further, the geometric mean of the sky and airmass auto-correlation functions is nearly identical to their cross-correlation function. This suggests that the two fields are nearly fully co-variant in terms of the information they provide on the large-scale clustering of LGs. One might expect that the two would be related, as the airmass and, in general, the sky background are higher closer to the horizon. (The sky background should also depend on, e.g., the phase of and proximity to the Moon and the azimuthal angle.) However, it is only the large scale patterns that are similar --- the local effects on the density field (as displayed in Fig. \ref{fig:ntarall}) clearly differ.

Finally, we use the results of \cite{Sch10}, who found color offsets (caused by some combination of errors in the Galactic dust map and/or photometric calibration errors) for SDSS data based on the blue tip of the stellar locus, to make a map of the offset in $d_{\perp}$. We note that $\sim$15\% of the imaging in the South was not available at the time these maps were made and that one may expect the color of the blue tip to vary with the average metallicity of the stars that are used (which will vary as a function of position in the Galaxy). Regardless, any fluctuations we find may be important. We test against the implied offset in $d_{\perp}$, as small changes in $d_{\perp} < 0.55$ cut have a large effect on the number of objects we select into our sample (see Section \ref{sec:NS}). We find there to be slight excess in the number of LGs we find at both low and high values of the offset (displayed in the upper-right panel of Fig. \ref{fig:ntarall}). The auto-correlation of the $d_{\perp}$ offset is significant and $\sim1/10^{th}$ that of $A_r$, but its cross-correlation function with the LGs is consistent with zero at nearly all of the scales we measure (both are displayed using magenta open circles in Fig. \ref{fig:wsys}).
\begin{figure}
\includegraphics[width=84mm]{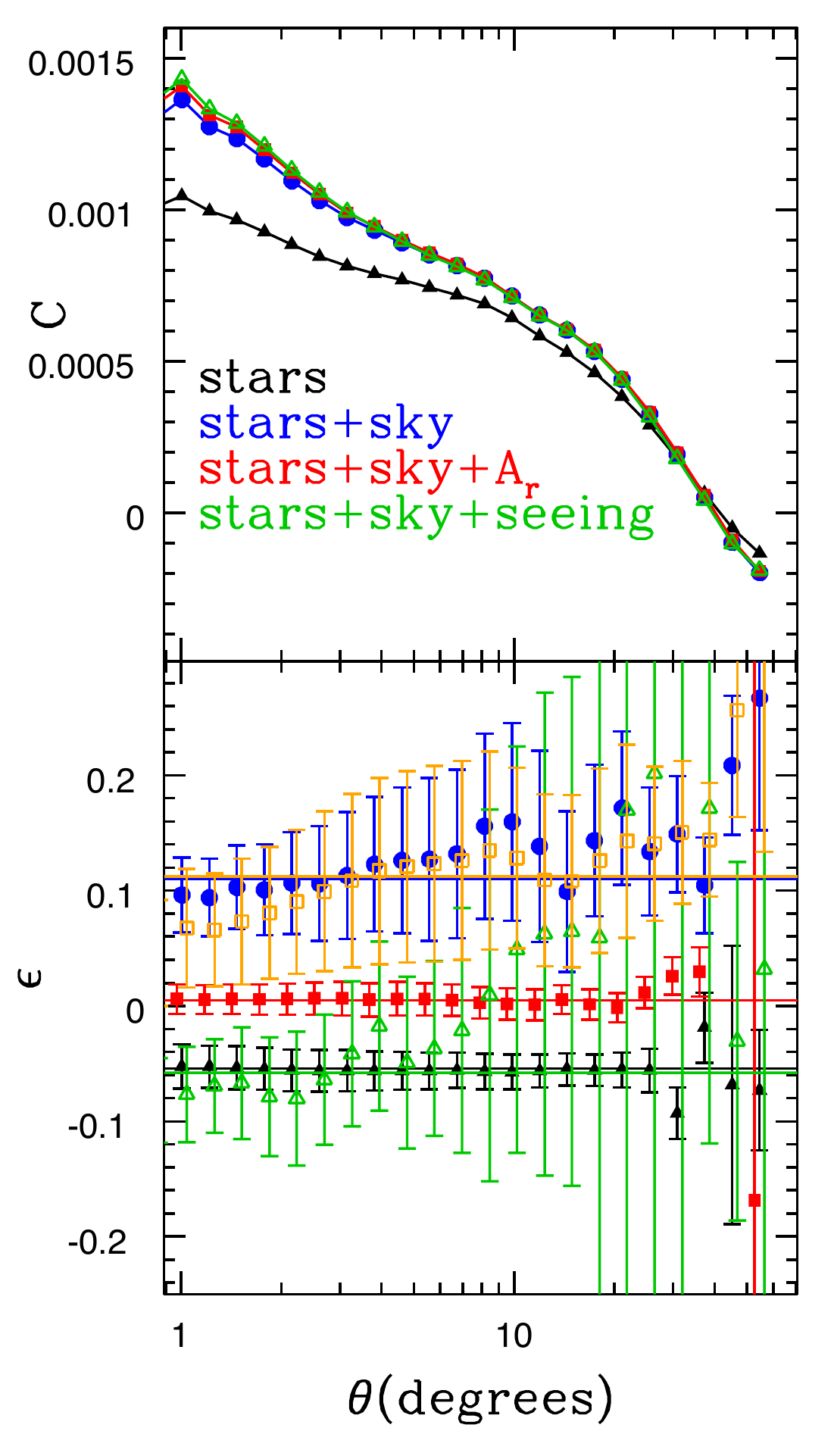}
\caption{Top panel: The correction, $C$, calculated as described by Eqs. \ref{eq:acc} and \ref{eq:ccc}, to the auto-correlation of the full LG catalog, when considering: stars only (black triangles); stars and sky background (blue circles); stars, sky background, and Galactic extinction in the $r$-band ($A_r$; red squares); and stars, sky background, and seeing (green open triangles). Bottom panel: points with error-bars (calculated by propagating the jack-knife errors on the auto- and cross-correlation functions) display the value of $\epsilon$ (see Section \ref{sec:wcorr}) for stars (black triangles), sky background (blue circles), extinction (red squares), and seeing (open green triangles), and airmass (open orange squares). The solid lines of corresponding color represent the best-fit constant value of $\epsilon$ for each respective systematic.}
\label{fig:wcorr}
\end{figure}

\subsection{Eliminating Systematic Errors}
\label{sec:csys}

We have investigated six potential sources of systematic errors (foreground stars, Galactic extinction, seeing, sky background, airmass, and photometric offsets) and we find different fluctuations in LG density associated with each. The auto-correlation functions of these potential systematics and their cross-correlation functions with the LGs suggest that stars have the greatest potential to cause systematic deviations in the measured clustering, and that we may have to worry about sky background fluctuations as well.
\begin{figure*}
\begin{minipage}{7in}
\includegraphics[width=7in]{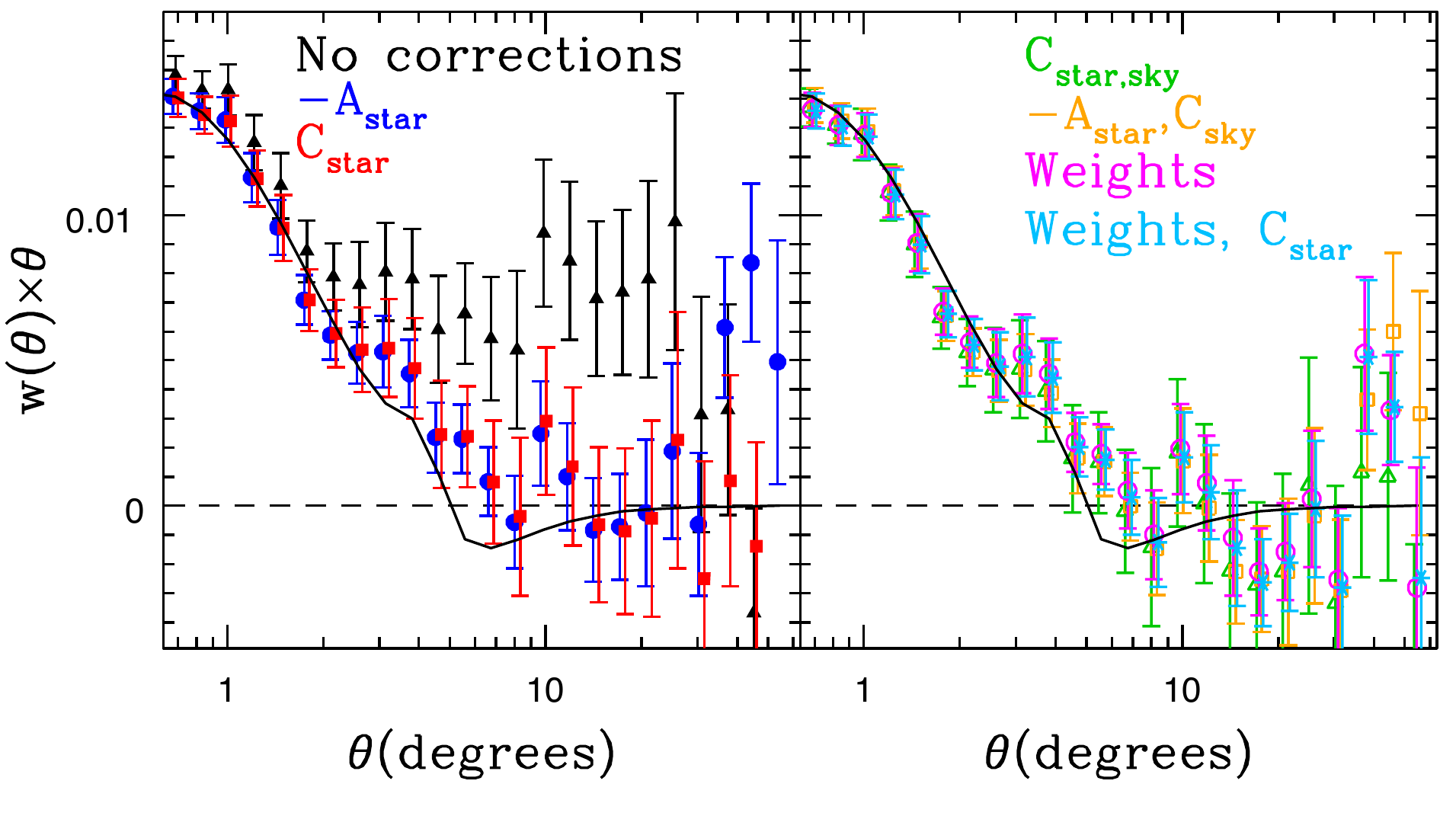}
\caption{The measured angular auto-correlation, $w(\theta)$, measured by weighting each LG by $p_{sg}$, multiplied by $\theta$ for our full sample of LGs. In the left-hand panel, the black triangles show the measurements with no corrections and the fiducial mask, the red squares mark the measurement when we correct for stars (as described in Section \ref{sec:wcorr}), and the blue circles display the measurement when we make the $A_{star}$ correction. In right-hand panel, the open green triangles display the measurement we obtain when we correct for stars and sky background ($C_{star,sky}$), the open orange squares display the results when we make the $A_{star}$ correction and additionally correct for sky background ($-A_{star},C_{sky}$), the open orange squares show the $w(\theta)$ we measure when we apply iterative weights to the LG density field (Weights), and the cyan stars show the result when we apply a $C_{star}$ correction in addition the weights. The solid line displays the model $w(\theta)$ for our assumed cosmology and $b=2$.}
\label{fig:wtst}
\end{minipage}
\end{figure*}

There are at least three procedures one can use to correct for these sources of potential systematic error. The first is to mask area of the sky based on the value of the observational parameter. For instance, we have already masked areas with $E(B-V) > 0.08$ and seeing $> 2.^{\prime\prime}0$. As we described in Section \ref{sec:fst}, masking may be effective for removing the effects of foreground stars. However, for the other potential systematics, there will remain fluctuations in the number density of LGs no matter the value of the cut we make on the systematic. 

A second option is to find the combination of weights one can apply to remove the fluctuations one finds in the LG number density. This application is straightforward in the case that the effects are uncorrelated --- each galaxy would just be weighted by the reciprocal of the function plotted in the bottom panels of Fig. \ref{fig:ntarall}. However, we find significant correlation between the effects, making this process non-trivial. 

The correlation between possible systematics can be accounted for by iteratively applying the weights. For instance, one may find the weights based on stellar density, $W_{star}(n_{star})$, by taking the reciprocal of what is plotted in red squares in Fig. \ref{fig:ntarall} and then find $\bar{n}/\bar{n}_{tot}(A_r)$ while applying $W_{star}$ to the density field, the reciprocal of which is (the independent) $W_A(A_r)$. One may then proceed likewise through all potential systematics. The disadvantage to this approach is that it assumes each effect is fully separable (i.e., that we can express $W[n_{star},A_r] = X[n_{star}]Y[A_r]$) and in the (realistic) case that this is not 100\% true, the order with which the weights are determined will matter. The advantage is that this method does not require a linear relationship between the potential systematic and the LG density fields and that it is straightforward to apply to as many potential systematics as can be identified. The resulting $\bar{n}/\bar{n}_{tot}(sys)$ relationships when we determine the weights iteratively, in the order $n_{star}$, airmass, seeing, $A_r$, $d_{\perp}$ offset, sky are displayed in orange in Fig. \ref{fig:ntarall}. Only for $n_{star}$ does the effect of applying subsequent weights cause a significant deviation from $\bar{n}/\bar{n}_{tot}(sys) = 1$. From hereon, we will refer to the application of these weights as the ``Weights'' method.

A third option is to use the cross-correlation technique described in Section \ref{sec:wcorr} to estimate and eliminate the spurious signal imparted by any number of observational parameters. We will refer to this method as the ``correction'' technique. The top panel of Fig. \ref{fig:wcorr} displays the magnitude of the correction, calculated as described in Section \ref{sec:wcorr}, for different combinations of observational parameters. The correction for the stars alone (black triangles) decreases from $\sim 10^{-3}$ to $\sim 4\times10^{-4}$ from $1^{\rm o}$ to $20^{\rm o}$. Additionally correcting for sky background (blue circles) marginally increases the correction at large scales but increases the correction by $\sim$30\% at $1^{\rm o}$. Adding seeing (red squares) or $A_r$ (green stars) corrections has only a marginal effect, which is most notable at small scales.

In the bottom panel of Fig. \ref{fig:wcorr}, the value of $\epsilon$ (as defined in Section \ref{sec:wcorr}) is displayed as a function of angle for stars (black triangles), sky background (blue circles), $A_r$ (red squares), seeing (open green triangles), and airmass (open orange squares). The correction we apply requires that $\epsilon$ be constant. The solid lines display the best-fit (constant) value of $\epsilon$. As can be seen, in every case a constant value of $\epsilon$ is well within the error-bars, suggesting no need for higher-order corrections. The values of $\epsilon$ are slightly more constant for sky background than for airmass, and we have found that they both trace the same large-scale clustering pattern. For this reason, we use only the sky background when calculating corrections.

Unless we note otherwise, in all cases we measure the angular auto-correlation function of LGs, $w(\theta)$, by weighting each object by the value of $p_{sg}$. This means that instead of counting each LG equally, a LG is counted as $p_{sg}$, and thus, at large smoothing scales, the estimated over-density of LGs should be the true (observed) over-density of LGs. In principle, this should remove the contamination of stars, leaving only their systematic masking effect. We display this $p_{sg}$ weighted measurement of $w(\theta)$ with no corrections, multiplied by $\theta$, for all of the LGs within our (masked) imaging area with black triangles in the left-hand panel of Fig. \ref{fig:wtst}. The amplitudes at large scales are quite large, given that generic models predict $w(\theta) \sim 0$ for $\theta > 5.0$. 

Subtracting $A_{star}$ for each star within a pixel (blue circles in the left panel of Fig. \ref{fig:wtst}) significantly reduces the large scale amplitudes of $w(\theta)$. Notably, this result is virtually identical to what we obtain when we do not subtract $A_{star}$ but instead apply the correction technique when accounting for stars, as displayed with red squares in the left-hand panel of Fig. \ref{fig:wtst} (the magnitude of this correction is displayed with black triangles in Fig. \ref{fig:wcorr}). This suggests that either method can be used, and the approach one uses should depend on its ease with respect to the task at hand. Notably, the jack-knife errors are smaller when we subtract $A_{star}$. 

The LG $w(\theta)$ with the combined correction for stars and sky background ($C_{star,sky}$) is displayed by the open green triangles in the right-hand panel of Fig. \ref{fig:wtst} --- including the sky background correction produces a small but noticeable change (the measurements around 3$^{\rm o}$ are closer to the black line), which is almost identical to subtracting $A_{star}$ and correcting for sky background at scales less than 30$^{\rm o}$ (orange open squares; $A_{star},C_{sky}$). Including additional corrections for seeing and Galactic extinction produces no discernible change in the LG $w(\theta)$. Applying the Weight method to the LG density field yields the measurements displayed with open magenta circles in the right hand panel. The results are very similar to the other correction techniques at scales less than 30$^{\rm o}$, but the measurements are slightly larger at scales greater than 2$^{\rm o}$ than either $A_{star}, C_{sky}$ or $C_{star,sky}$. The data displayed in orange in Fig. \ref{fig:ntarall} suggest that the Weight method may leave a residual dependence on stellar density. If we subtract the correction for stellar density we find by cross-correlating the weighted LG field with the stars, the resulting effect on the $w(\theta)$ (displayed with cyan stars in the right-hand panel of Fig. \ref{fig:wtst}) measurements is minor. The disagreement between the results at $\theta > 30^{\rm o}$ suggests that significant systematic errors remain on the measurements at these scales.

The black curve plotted in Fig. \ref{fig:wtst} displays the expected clustering for our fiducial $\Lambda$CDM cosmology and a bias of 2.0 (calculated as described in Section \ref{sec:th}). This model appears generally consistent with all of the measurements at small scales, but at scales greater than 2$^{\rm o}$, the un-corrected measurements are significantly greater. We note that the feature in the model at $\sim3.5^{\rm o}$ is due to the baryon acoustic oscillations present in our fiducial $P(k)$. Including corrections for stars and sky background appears to make the measurements generally consistent with the assumed cosmological model, although all but one of the measurements remain larger than the model between $3^{\rm o}$ and 12$^{\rm o}$.

\subsection{North and South Galactic Caps}
\label{sec:NS}
\begin{figure}
\includegraphics[width=84mm]{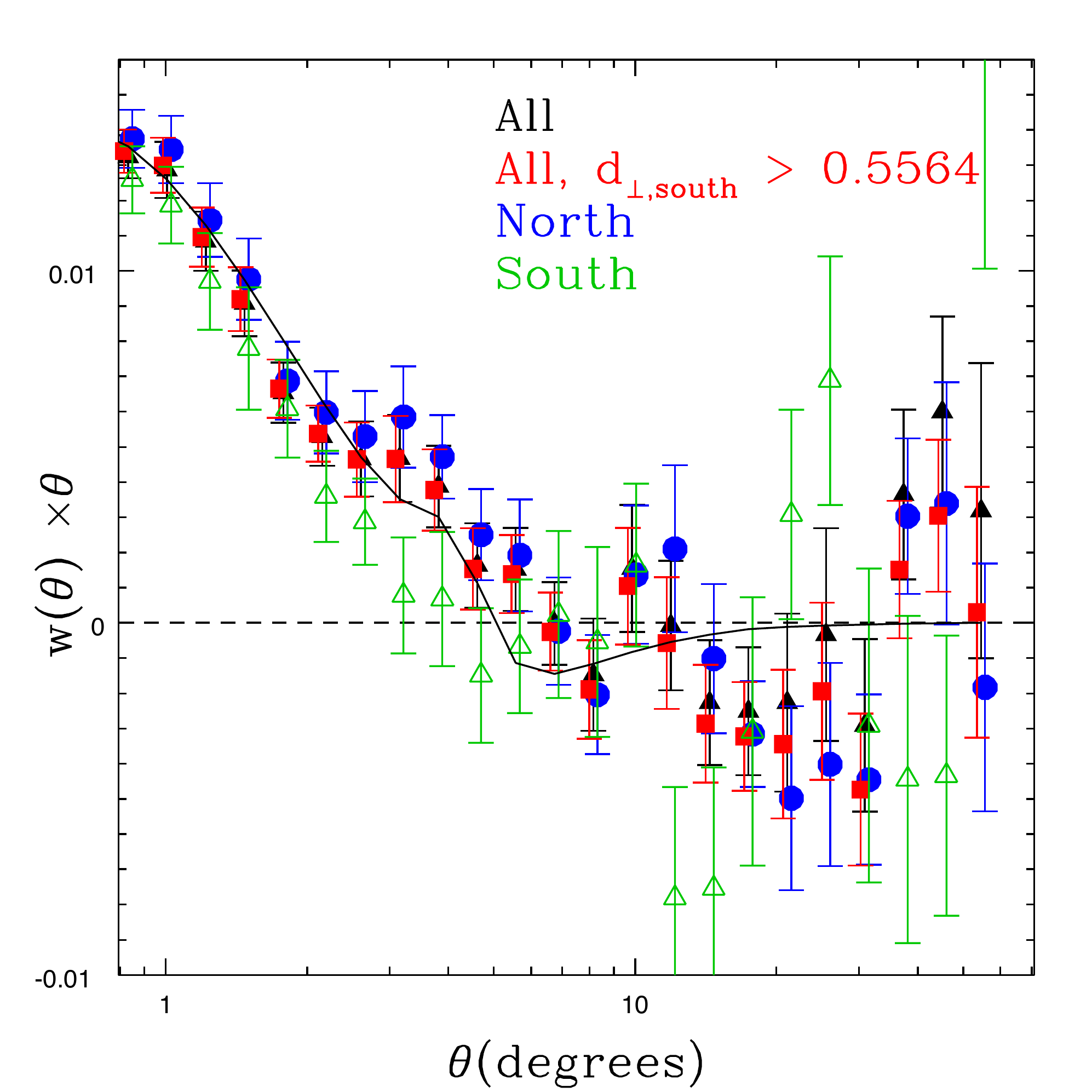}
\caption{The measured angular auto-correlation functions ($w[\theta]$) for the full LG sample (black triangles), LGs in the Northern Galactic Cap (blue circles), LGs in the Southern Galactic Cap (open green triangles), and the LG sample we obtain when we impose the cut $d_{\perp} > 0.5564$ for objects in the South Galactic Cap (red squares). In each case, we subtract an area $A_{star}$ for each star in each pixel and correct for sky brightness.}
\label{fig:wNS}
\end{figure}

The DR8 imaging data is separated into two distinct regions; in Galactic coordinates these regions can be separated into $b > 0^{\rm o}$ `North' and $b < 0 ^{\rm o}$ `South' (see Fig. \ref{fig:footprint}). The fact that these regions are spatially separated makes them more prone to calibration errors, e.g., one might expect uncertainty in the relative zero points of one or more bands, given the lack of continuous photometry connecting the two regions. \cite{Sch10} and \cite{Sch10b} have estimated the level of color offset between the North and South in the SDSS. \cite{Sch10b} attribute these differences to either calibration errors or errors in the Galactic extinction corrections (or a combination of both). Using results of the `spectrum' based method (which is less sensitive to changes in the metallicity of stars than the `blue tip' method) listed in the second row of Table 6 of \cite{Sch10b}, one can infer that the values of $d_{\perp}$ (a combination of $g-r$ and $r-i$ colors defined by Eq. \ref{eq:dp}) that we calculate are offset by 0.0064 magnitudes between the North and South (the `blue tip' method yields a similar offset of 0.0045 magnitudes). These results suggest that, assuming the values in the North are the true values, we should subtract 0.0064 from the values we calculate in the South to obtain a better estimate of their $d_{\perp}$ values. 

The inferred offset in $d_{\perp}$ between the North and South appears small, and given the difference between the `spectrum' and `blue tip' based methods, the uncertainty on the correction may be relatively large. However, it is instructive to determine the effect a 0.0064 magnitude offset in $d_{\perp}$ would have on our sample. If we change the cut such that we accept only objects with $d_{\perp} > 0.5564$ in the South (while using the fiducial cut in the North), we remove 5,172 objects from our sample. This reduces the number density in the South to 108.7 deg$^{-2}$ --- which is still 1.5\% greater than the number density in the North. If we weight each object by $p_{sg}$ when calculating the number densities (which should provide a better estimate of the true number density of galaxies) the number density in the South decreases to 103.2 deg$^{-2}$ and in the North it becomes 103.1 deg$^{-2}$. Any uncertainty in the corrections we have made is almost certainly larger than this 0.1 deg$^{-2}$ difference in number density. The $\bar{n}/\bar{n}_{\rm tot}(sys)$ relationships we obtain when we use the $d_{\perp} > 0.5564$ cut in the South, subtract $A_{star}$, and weight by $p_{sg}$ are displayed in green in Fig. \ref{fig:ntarall}. This does not cause major changes, but the relationships with airmass, $A_r$, and $d_{\perp}$ offset all become closer to one.

Given the color offsets and the change in number density they imply, we might expect differences in the clustering of the LGs in the North and South. Fig. \ref{fig:wNS} displays the $w(\theta)$ we measure when we split the LG sample into North (blue circles) and South (open green triangles) samples and apply the $A_{star},C_{\rm sky}$ correction. Indeed, the clustering is different in the two regions, as the measurements in the South are smaller than those in the North. However, at scales less than 30$^{\rm o}$, the $w(\theta)$ of full sample (black triangles) simply appears to be the weighted average of the two samples. At the largest scales ($\theta > 30^{\rm o}$), this is not the case, suggesting that the color offset may cause a systematic effect on the measurements at the largest scales (we note that the sky background correction may have minimized this potential systematic effect, confining it to these large scales).

We have found that using the cut $d_{\perp} > 0.5564$ in the South removes the asymmetry in the number density of LGs in the North and South. We therefore measure $w(\theta)$ of the LG sample after making this cut, which is plotted with red squares in Fig. \ref{fig:wNS}. The result is quite similar to the result obtained using the fiducial cuts (black triangles), but the amplitudes at the largest scales are reduced and appear to be closer to the weighted average of the North and South. Interestingly, the size of the sky brightness correction depends strongly on our particular treatment of the North and South. We find $\epsilon_{\rm sky} = 0.113$ for the full sample, $\epsilon_{\rm sky} = 0.068$ when we use the $d_{\perp} > 0.5564$ cut in the South, $\epsilon_{\rm sky} = 0.027$ in the North sample, and that $\epsilon_{\rm sky} = 0.18$ for the South sample. This suggests that the systematic effect of sky brightness predominantly a feature of the Southern imaging.

\subsection{Summary of Angular Fluctuations}
The results presented throughout this section suggest that stars cause major systematic errors on the clustering of SDSS DR8 LGs, and sky brightness may also cause significant errors. We have investigated variations in number density caused by Galactic extinction, seeing, airmass, and color offsets, but found them to have minor effects. Perhaps most importantly, we have identified two separate ways to correct for systematic variations in the number density of galaxies caused by any potential systematic that can be quantified and turned into a map on the sky, and for the stars we have identified three separate ways to correct for their systematic effects.

We note that other catalogs constructed from other imaging surveys, other SDSS data, or even subsets of the LG data will not necessarily display the same relationships we have found in this section. The tests we have performed must be repeated for any sample one uses to measure clustering. We further note that the systematics we investigate are by no means a complete list --- there are likely to be effects we have not thought of. 

\section{Constructing the Photometric Redshift Catalog}
\label{sec:spec}
We measure photometric redshifts for our LG sample using the artificial neural network based photometric redshift estimator ANNz \citep{F03}. ANNz has been proven to yield accurate and precise $z_{phot}$ estimates when the training sample is representative of the full data set (see, e.g., \citealt{megaz}). The results of \cite{ab08} and \cite{thomas10} suggest that neural-network based photometric redshift estimators (such as ANNz) are the most accurate in this specific situation. Our training sample consists of 112,778 BOSS CMASS objects with spectroscopic redshifts. This large training sample provides unprecedented ability to ensure the training sample is representative of our imaging data, while accounting for fluctuations in observing conditions. 

\begin{figure}
\includegraphics[width=84mm]{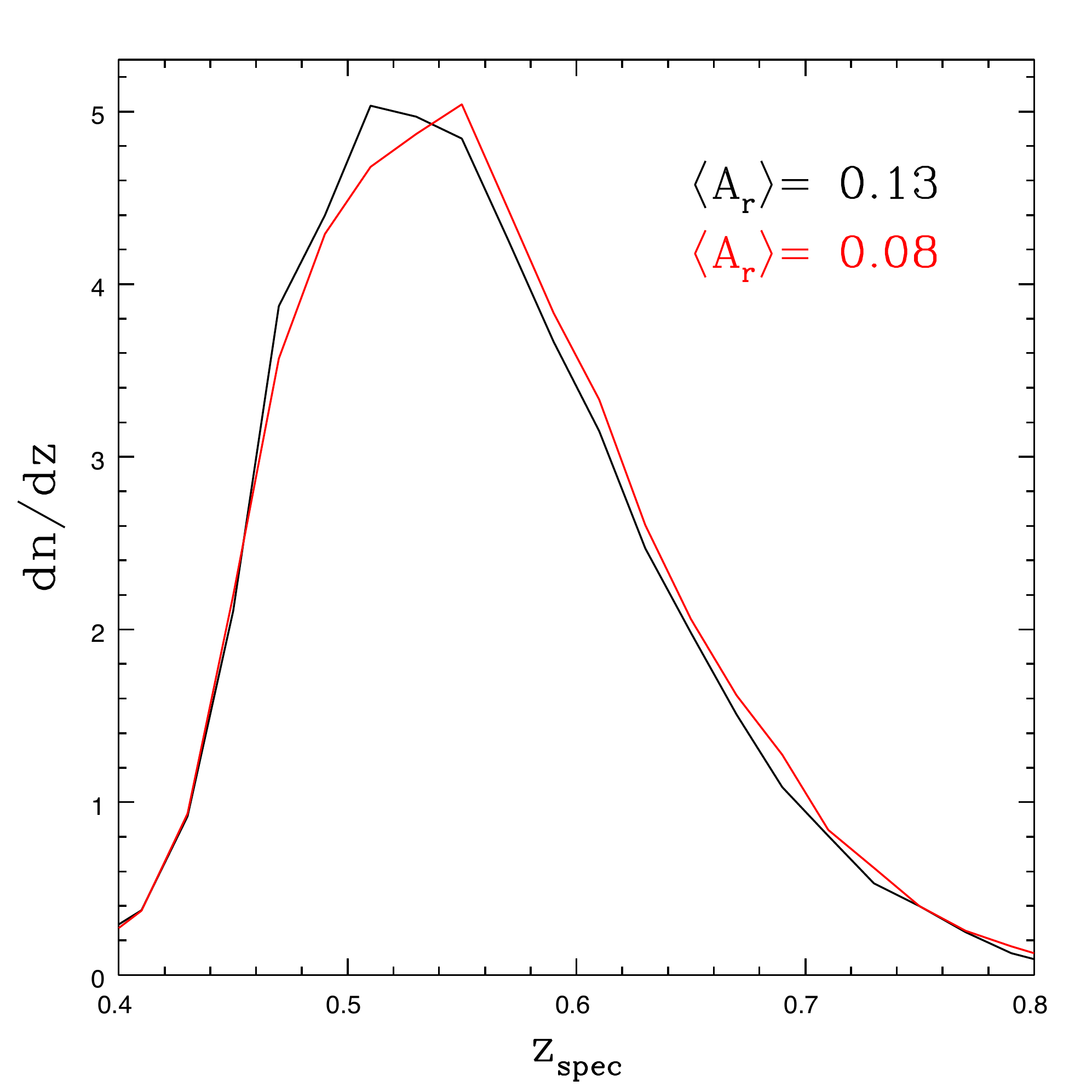}
\caption{The normalized redshift distributions of BOSS CMASS spectra when splitting into regions with $r$-band Galactic extinction, $\langle A_r \rangle$, = 0.08 (red) and 0.13 (black)}
\label{fig:nzhl4pan}
\end{figure}

Fig. \ref{fig:nzhl4pan} displays the normalized redshift distributions for data with $\langle A_r\rangle = 0.08$ and $\langle A_r \rangle = 0.13$ (which splits the sample $\sim$ in half). There is a significant difference, as the objects in areas of the sky with low Galactic extinction (red) have a larger median redshift and more galaxies in the high redshift tail of the distribution. This result suggests that we may obtain better $z_{phot}$ estimates if we include the $A_r$ values in the training, which we can do because the training data cover the entire range of Galactic extinction found in our full sample. This finding implies that Galactic extinction may be an important systematic when the clustering of the BOSS spectroscopic sample is analyzed.

We have also found fluctuations in the spectroscopic redshift distributions with seeing and sky background; these fluctuations will be studied in detail in \cite{Pspec}. Despite the fact that the training consists of over 100,000 objects, we do not find that it adequately covers the range in sky background or seeing which would be required to include these parameters in the photometric redshift training. Instead, we repeat the tests we performed on the whole sample in Section \ref{sec:tar} on samples in $z_{phot}$ slices in Section \ref{sec:wmeas}.

\subsection{Photometric Redshift Training}
\begin{figure*}
\begin{minipage}{7in}
\includegraphics[width=7in]{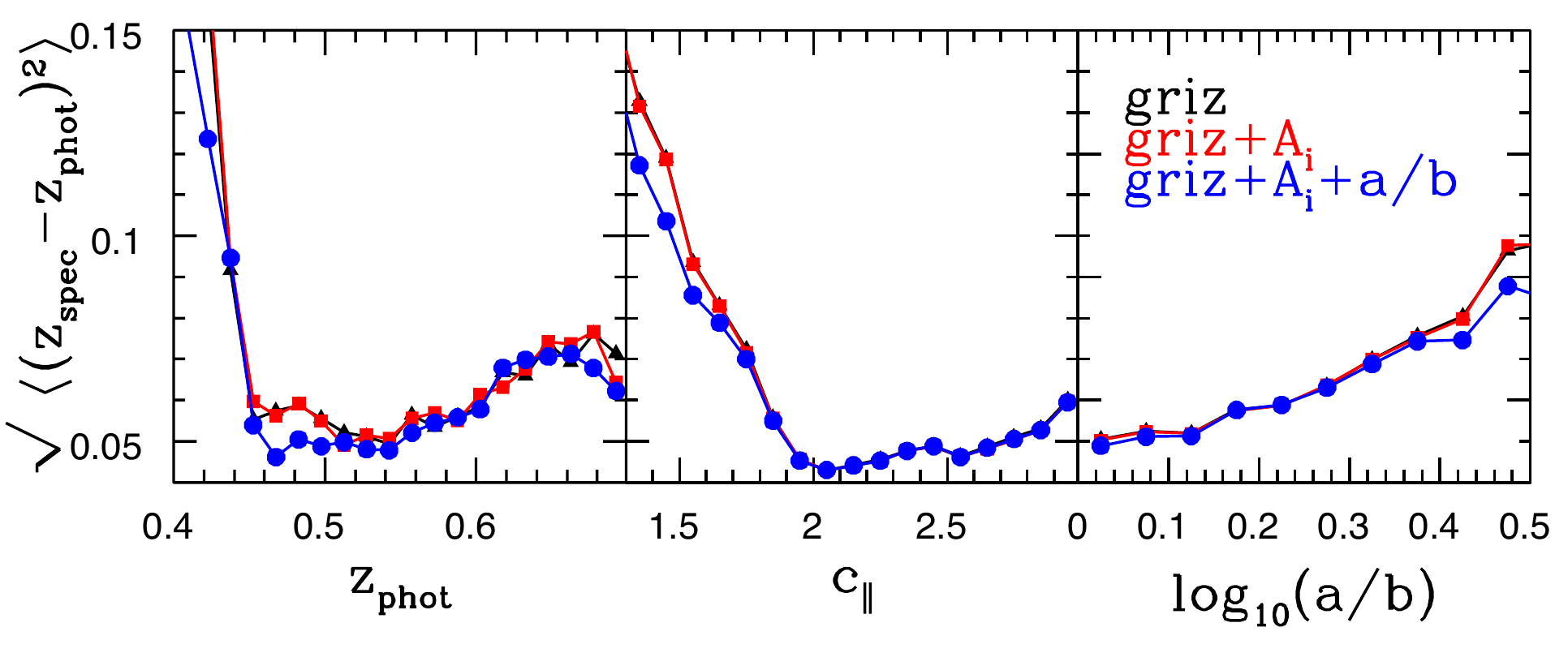}
\caption{The rms between the $z_{phot}$ estimate and the spectroscopic redshift in the BOSS testing set, as a function of the $z_{phot}$ estimate (left panel), $c_{\parallel}$ (middle panel), and the axis ratio for ellipse corresponding to the exponential fit to the $i$-band profile (right). We display results for the three separate spectroscopic redshift samples we use to train ANNz: i) when we train using only the $g,r,i,z$ (de-reddened) model magnitudes (black triangles), ii) when we also include the Galactic extinction in the $i$-band ($A_i$; $A_i=2.086E(B-V)$; red squares), and, iii) when we also include the axis ratio for the exponential fit to the $i$-band profile ($a/b$; blue circles).}
\label{fig:pzrmsall}
\end{minipage}
\end{figure*}

We train ANNz to estimate photometric redshifts for our LGs using similar methods as \cite{megaz}. We randomly divide our sample of CMASS spectra (keeping only objects spectroscopically confirmed to be galaxies) such that one quarter of the objects are designated as a training set, a separate one quarter as a validation set, and the remaining half of the objects as a testing set (as is done in \citealt{megaz}).  We use the de-reddened model magnitudes in the $g,r,i,z$ SDSS imaging bands, and their errors. We do not use the $u$-band because the results of \cite{Sch10} suggest there may be significant variations in the true $u-g$ color over the SDSS imaging area, and the $u$-band only significantly aids $z_{phot}$ estimation for the bluest galaxies in our sample. 

We test three different training samples that use the following input parameters:
 \begin{enumerate} \item Only the $g,r,i,z$ (de-reddened) model magnitudes and their errors \item Including the Galactic extinction in the $i$-band, $A_i$ in addition to the model magnitudes \item Including the ratio of major to minor axes\footnote{The data selected from the CAS are $AB_{exp}$, which are actually $b/a$ ratios}, $a/b$, of the ellipse corresponding to the best-fit exponential profile in the $i$-band in addition to the magnitude and $A_i$ information \end{enumerate} 

In case (i), the rms difference between $z_{spec}$ and $z_{phot}$ is 0.0610 for the full sample; this reduces to 0.0512 for galaxies with $c_{\parallel} > 1.6$. We find that 92\% of the LGs pass this $c_{\parallel} > 1.6$ restriction which was used by previous studies, such as \cite{megaz}. Such a restriction ensures a strong 4000${\rm \AA}$ break for galaxies at our target redshifts ($0.4 < z < 0.7$), and \cite{Masters11} find that this cut removes most of the galaxies that would be morphologically classified as late-type from the CMASS sample. For case (ii), we find that including the $A_i$ information produces insignificant improvements, as the rms values become 0.0609 and 0.0511, respectively. Case (iii) significantly improves the rms for objects with $c_{\parallel} < 1.6$, as the overall dispersion decreases to 0.0585 and we find a slight improvement for the $c_{\parallel} > 1.6$ galaxies, as the rms decreases to 0.0506.

As the dispersion values suggest, there is a strong correlation between the value of  $c_{\parallel}$ and the accuracy with which we can estimate $z_{phot}$. We display the relationship between the rms dispersion in our testing set versus $c_{\parallel}$ in the middle panel of Fig. \ref{fig:pzrmsall} for our three different training samples. The obtained relationship is extremely similar whether we include $A_i$ (case ii, red squares) or we do not (case i, black triangles). Including $a/b$ (case iii) reduces the rms for objects with $c_{\parallel} < 1.8$. However, all three cases show that the most accurate $z_{phot}$ estimates are obtained when $c_{\parallel} > 1.8$.

We also find a significant correlation between the accuracy of the $z_{phot}$ estimates and $a/b$. \cite{Masters11} has discovered that $\sim$5\% of CMASS objects are edge-on spiral galaxies where effects of dust obscuration are likely to be significant. Logically, this likelihood is correlated with the axis ratio. These dust-obscured spirals tend to be at lower redshifts than the majority of CMASS objects (while having similar colours, see \citealt{Y11} for a full study of the effects of inclination on photometric redshift estimates), and thus the accuracy of the $z_{phot}$ estimate and $a/b$ are related. We present this relation in the right panel of Fig. \ref{fig:pzrmsall}, which shows that it is a smooth function of log$_{10}(a/b)$ for each of our three training samples. Including $a/b$ in the training improves the $z_{phot}$ accuracy for the largest values of $a/b$. The values of $a/b$ are correlated with $c_{\parallel}$, since it is disk galaxies (which are generally bluer galaxies in $c_{\parallel}$) that have the highest values of $a/b$. However, we find similar relationships (though not as strong) when this correlation is accounted for.

The dispersion is also correlated with the estimated $z_{phot}$, as illustrated in the left panel of Fig. \ref{fig:pzrmsall}. Each of the three training cases result in similar relationships. Including $a/b$ (case iii, blue circles) makes the largest difference for $z_{phot}$ estimates between 0.45 and 0.6. We see only minor differences between cases i (model magnitudes only, black triangles) and ii (including $A_i$, red squares).

The ANNz output includes a photometric redshift error estimate, which we denote $\sigma_{ze}$. These reported errors are correlated to the actual dispersion in  $z_{phot}$ vs. $z_{spec}$, but they underestimate it by a factor of $\sim$66\% (as can be determined by comparing the average value of $\sigma_{ze}$ and $\sqrt{\langle (z_{phot} - z_{spec})^2\rangle}$ for any particular testing set). These estimated errors do not recover the trends we discover between the rms and $c_{\parallel}$ and $a/b$ (which are displayed in the middle and right panels of Fig. \ref{fig:pzrmsall}). Thus, the true uncertainty on any individual $z_{phot}$ estimate is a linear combination of the estimated error, $c_{\parallel}$, and $a/b$. In section \ref{sec:edndz}, we describe how we combine this information in order to estimate the redshift distributions of the photometric redshift samples we use.

\begin{figure}
\includegraphics[width=84mm]{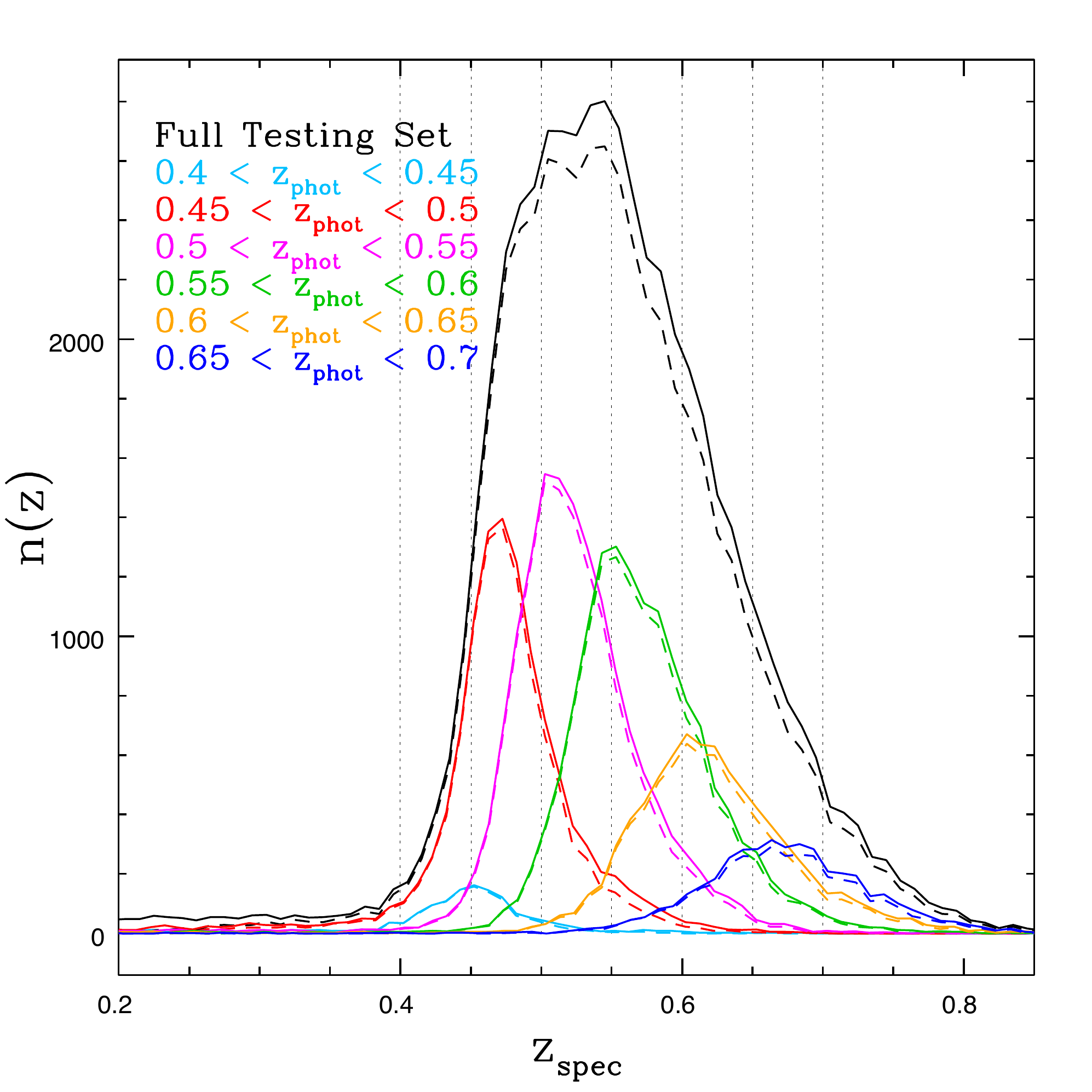}
\caption{The redshift distributions of spectroscopic galaxies in our testing set, for different $z_{phot}$ selections. The dashed line represent galaxies with $c_{\parallel} > 1.6$. The vertical dotted lines delineate the $z_{phot}$ bounds of the six $\Delta z_{phot} = 0.05$ redshift slicess that are displayed. }
\label{fig:nzall}
\end{figure}

Fig. \ref{fig:nzall} presents the overall redshift distribution of spectroscopic galaxies in our testing set (solid black line). The colored lines display the spectroscopic redshift distributions of testing set galaxies in slices of width $\Delta z_{phot} = 0.05$ from 0.4 to 0.7, when we estimate $z_{phot}$ using case iii (including $A_i$ and $a/b$ in the training sample). In all cases, the dashed lines represent galaxies with $c_{\parallel} > 1.6$.  These distributions suggest that, if one wishes to use slices of width $\Delta z_{phot} = 0.05$, the bins $0.45 < z_{phot} < 0.5$, $0.5 < z_{phot} < 0.55$, $0.55 < z_{phot} < 0.6$, and $0.6 < z_{phot} < 0.65$ contain most of the information, as the bins $0.4 < z_{phot} < 0.45$ and $0.65 < z_{phot} < 0.7$ have their true redshifts almost entirely within the adjacent $z_{phot}$ slice.

\subsection{Photometric Redshift Catalog}
\label{sec:phz}
We construct a photometric redshift catalog using the objects selected as described in Section \ref{sec:data}, and using the ANNz training which includes both Galactic extinction and axis ratios (case iii). We did not find any significant difference in the accuracy of the $z_{phot}$ estimates when we added $A_i$ information to the training (case ii). However, we do find a large difference in the full $z_{phot}$ distributions. In particular, we find that the asymmetry between the North and South increases as a function of $z_{phot}$. For $0.6 < z_{phot} < 0.65$, $c_{\parallel} > 1.6$, and weighting by $p_{sg}$, we find a 7.5\% larger number density of LGs in the South when we do not include $A_i$ values in the training, and only a 4.9\% larger number density when we do include $A_i$. 

\begin{figure}
\includegraphics[width=84mm]{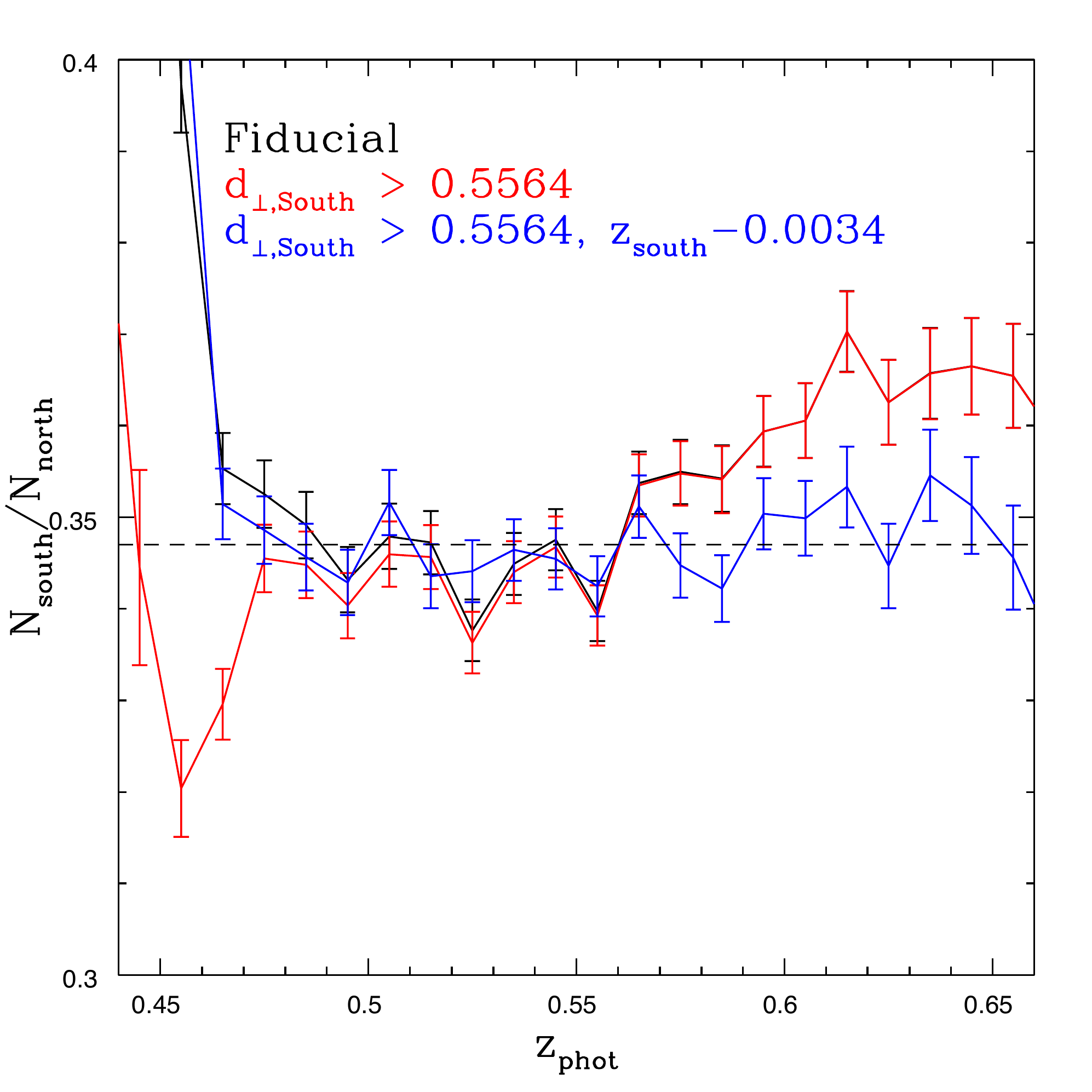}
\caption{The ratio between the number of objects, when weighting by $p_{sg}$, in the South and North, as a function of $z_{phot}$ for our fiducial sample (black), when we apply the cut $d_{\perp} > 0.5564$ to objects in the South (red), and when we apply the cut $d_{\perp} > 0.5564$ and subtract 0.0034 from every $z_{phot}$ for objects in the South (blue). The dashed black line displays the ratio of area in the South and North. Errors are Poisson.}
\label{fig:nznsrat}
\end{figure}

In Fig. \ref{fig:nznsrat} we display the ratio of the number of objects in the South to the number of objects in the North as a function of the $z_{phot}$ estimate, in black (we note that at high redshift, the red curve overlaps the black curve). The dashed black line shows the ratio of the area in South to the area in the North (0.347). For $z_{phot} > 0.55$ and $z_{phot} < 0.46$, there is a significant excess in the number of objects in the South. When apply the $d_{\perp} > 0.5564$ cut to objects in the South, there is a significant decrease in the number objects in the South with $z_{phot} < 0.46$, however, we find almost no change at larger photometric redshifts. This is due to the fact that objects that are assigned larger $z_{phot}$ have larger $d_{\perp}$ values. In fact, we find linear relationship between the average $z_{phot}$ and $d_{\perp}$, given by $z_{phot} = 0.53d_{\perp}$. Inserting the $\Delta d_{\perp} = 0.0064$ offset in for objects in the South (as suggested by \citealt{Sch10b}) yields a bias of $\Delta z = 0.0034$ for objects in the South. When we subtract 0.0034 from each $z_{phot}$ in the South, we find that the ratio between the number of objects in South and North (the blue curve in Fig. \ref{fig:nznsrat}) becomes nearly constant as a function of $z_{phot}$. As we found in Section \ref{sec:NS}, we find that assuming a difference in $d_{\perp}$ of 0.0064 (and its full consequences), removes the asymmetry between the distribution of objects in the North and South.

Fig. \ref{fig:nzall} implies that the majority of the cosmological information will be located within four $z_{phot}$ bins $0.45 < z_{phot} < 0.5$ (which we denote 1), $0.5 < z_{phot} < 0.55$ (which we denote 2), $0.55 < z_{phot} < 0.6$ (which we denote 3), and $0.6 < z_{phot} < 0.65$ (which we denote 4). The characteristics of each bin are summarised in Table \ref{tab:bins}. The training further suggests that we can only obtain accurate $z_{phot}$ estimates for objects with $c_{\parallel} > 1.6$, thus we also make this cut in each bin.

\begin{table}
\caption{The characteristics of the four photometric redshift ($z_{phot}$) bins we use, where $\sigma_{zo}$ refers to the value of $\sqrt{\langle (z_{phot} - z_{spec})^2\rangle}$ in the testing set and $\sigma_{zt}$ refers to the level of dispersion we infer for the full data set, using the methods described in Section \ref{sec:edndz}.} 
\begin{tabular}{lcccc}
\hline
\hline
bin & $z_{phot}$ range & $\sum_{i=1}^{i=N_{gal}} p_{sg}$ & $\sigma_{zo}$ & $\sigma_{zt}$\\
\hline
1 & $0.45 < z < 0.5$ & 214,971 & 0.0427 & 0.0431\\
2 &  $0.5 < z < 0.55$ & 258,736 & 0.0427 & 0.0442\\
3 & $0.55 < z < 0.6$ & 248,895 & 0.0501 & 0.0524\\
4 & $0.6 < z < 0.65$ & 150,319 & 0.0601 & 0.0633\\
\hline
\label{tab:bins}
\end{tabular}
\end{table}

\subsection{Estimating True Redshift Distributions}
\label{sec:edndz}
\begin{figure}
\includegraphics[width=84mm]{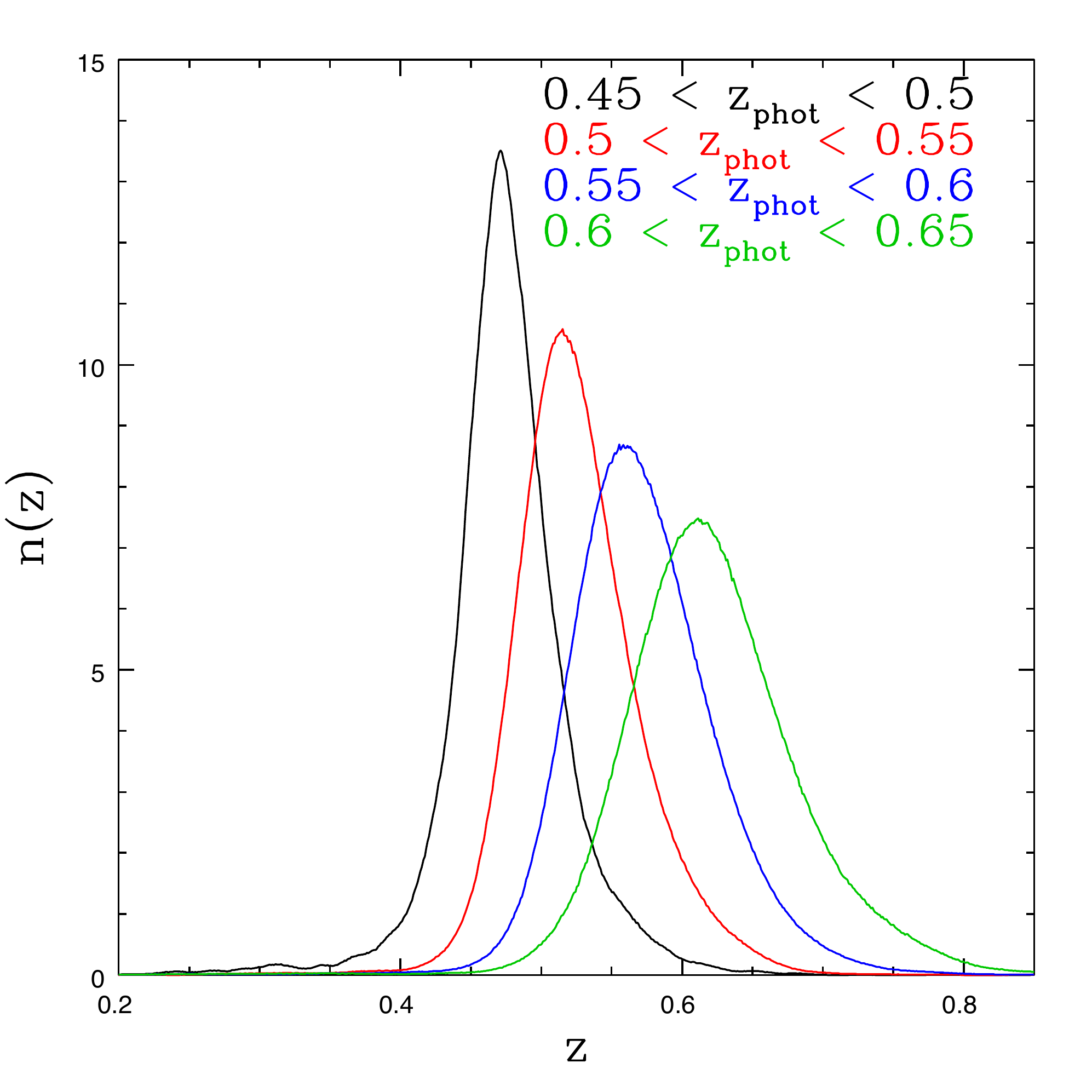}
\caption{The estimated (normalized) redshift distributions for the four labeled $z_{phot}$ slices, determined by sampling Gaussians around each spectroscopic LG in the testing data, as described in Section \ref{sec:edndz}. }
\label{fig:nzcg}
\end{figure}

To properly analyze any angular clustering measurement, one must know the true redshift distribution of the galaxies being used. This task is made relatively simple for the photometric redshift catalog we produce, as we expect the distributions to be similar to those of the training sample. However, the match is not perfect, and blindly assuming that the full catalog has the same distribution as the training sample would be folly --- differences between the catalogs must be accounted for. Based on our testing sample, we found that the actual dispersion between the photometric and spectroscopic redshift was well correlated to the error estimate, but also subject to the values of $c_{\parallel}$ and $a/b$. Thus, we can compare the distributions of photometric redshifts, error estimates, $c_{\parallel}$, and $a/b$ in the full data set to that of the testing set and use this information to estimate the true redshift distribution in any $z_{phot}$ slice.

For bin 1 ($0.45 < z_{phot} < 0.5$), the mean of the $z_{phot}$ error estimated by ANNz, $\bar{\sigma}_{ze}$, is 0.0224, while it is slightly lower, 0.0222, for the testing set (recall that these estimates underestimate the actual $\sqrt{\langle (z_{phot} - z_{spec})^2\rangle}$ by $\sim$66\%). The average $c_{\parallel}$ and $a/b$ of the full bin 1 and the testing set subset of bin 1 agree within $0.3$\%. We find the deviations in $c_{\parallel}$ are similarly small for the other three redshift bins. Overall this suggests that the true redshift distribution of bin 1 is slightly wider than the spectroscopic distribution, due to the fact that its $\bar{\sigma}_{ze}$ are 1\% larger than for the testing set. Thus, referring to $\sigma_{zt}$ as the true dispersion in the bin and $\sigma_{zo}$ as the dispersion in the testing set, we estimate $\sigma_{zt} = 0.0431$ given that $\sigma_{zo} = 0.0427$. In Table \ref{tab:bins}, we list the $\sigma_{zo}$ we measure from the testing set and the $\sigma_{zt}$ we estimate for each photometric redshift bin.

For bin 2 ($0.5 < z_{phot} < 0.55$), the differences are more substantial. The value of $\bar{\sigma}_{ze}$ is 3\% larger for the full catalog (0.0268 compared to 0.0260) and the average value of log$_{10}(a/b)$ is  2\% larger (0.170 compared to 0.166). Fig. \ref{fig:pzrmsall} suggests a linear relationship between the photometric redshift dispersion and log$_{10}(a/b)$ for $0 < {\rm log}_{10}(a/b) < 0.3$ that is $\sigma_{z}\sim 0.05 + 0.067{\rm log}_{10}(a/b)$ suggesting the overall error should be 3.5\% larger in bin 2 than for the testing set. The differences grow larger for bin 3 ($0.55 < z_{phot} < 0.6$): The value of $\bar{\sigma}_{ze}$ is 4\% larger (0.0307 compared to 0.0296) than the testing set and the average value of log$_{10}(a/b)$ is 5\% larger (0.180 compared to 0.172), suggesting the errors in bin 3 are 4.6\% larger than in the testing set. We find similar differences for bin 4 ($0.6 < z_{phot} < 0.65$: The values of $\bar{\sigma}_{ze}$ are 0.0334 and 0.0320 and the average log$_{10}(a/b)$ are 0.185 and 0.176, and we therefore expect the errors to be 5.4\% larger than in the testing set.

To correct for the differences between the testing set and full sample, we sample a Gaussian, for each spectroscopic redshift within the photometric redshift bin, of width such that the average dispersion in the bin increases to that which we expect. The dispersion we expect, $\sigma_{t}$, is related to the dispersion in the testing set, $\sigma_{zo}$, and the width of the Gaussian, $\sigma_d$, via
\begin{equation}
\sigma_{zt}^2 = \sigma_{zo}^2 + \sigma_d^2 .
\end{equation}
We are assuming we can relate $\sigma_{zt} = a\sigma_{zo}$, as described in the previous paragraph. Therefore
\begin{equation}
\sigma_d^2 = \sigma^2_{\rm zo}(a^2-1).
\end{equation}
For bin 4, $a = 1.054$ (since we determined the dispersion should be 5.4\% larger for the full sample than for the testing set) and $\sigma_{zo} = 0.06$, which yields $\sigma_d = 0.02$. We find that, with sufficient sampling, the resulting $n(z)$ are smooth up to steps of 0.001 in $\Delta z$. We display the $n(z)$ we obtain for our four bins in Fig. \ref{fig:nzcg}. In general, the displayed $n(z)$ are poorly fit by distributions such as a Gaussian or a Lorentzian (especially at the tails). Thus, we use the plotted distributions, which are binned in steps of width 0.001 in redshift, and interpolate between these points as needed in order to obtain the $n(z)$ used in Eq. \ref{eq:wnz}.

\begin{figure*}
\begin{minipage}{7in}
\centering
\includegraphics[width=180mm]{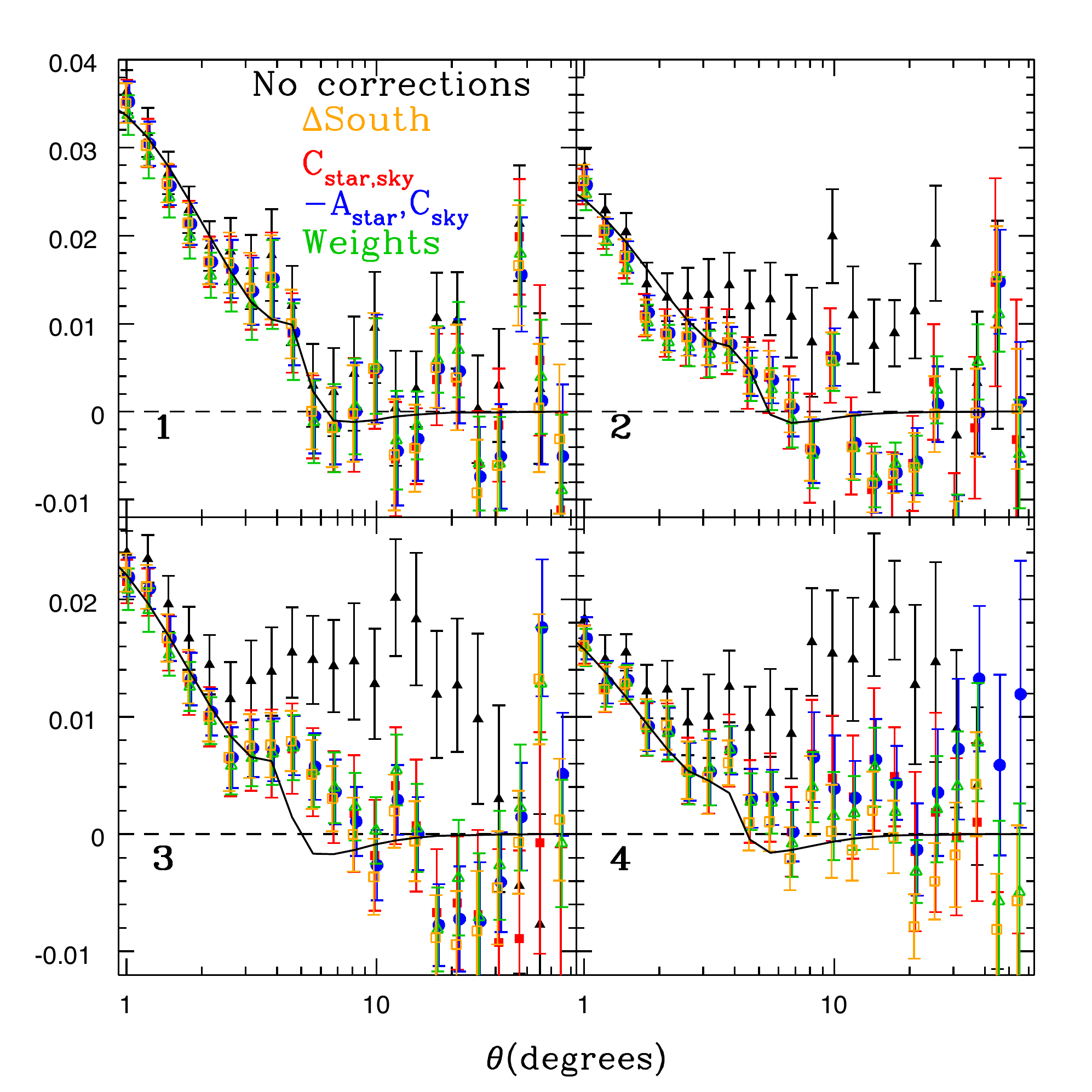}
\caption{The measured angular auto-correlation functions, $w(\theta)$, for our photometric redshift bins $0.45 < z_{phot} < 0.5$ (top left), $0.5 < z_{phot} < 0.55$ (top right), $0.55 < z_{phot} < 0.6$ (bottom left), and $0.6 < z_{phot} < 0.65$ (bottom right), when no corrections are applied (black triangles), when corrections for stars and sky background are applied in the manner described by Section \ref{sec:wcorr} (red squares; $C_{star,sky}$), when the effective area of stars, $A_{star}$, is removed from each pixel and a correction for sky background is applied (blue circles, $-A_{star},C_{sky}$), when iterative weights are applied to the LG density field used to calculate $w(\theta)$ (green open triangles), and when $-A_{star},C_{sky}$ is applied to the $w(\theta)$ of LGs selected such that $d_{\perp} > 0.5564$ and 0.0034 is subtracted from $z_{phot}$ for objects in the South (orange open squares; $\Delta$South).}
\label{fig:w4pansub}
\end{minipage}
\end{figure*}

\begin{table*}
\begin{minipage}{7in}
\caption{The minimum $\chi^2$ per degree of freedom ($\chi^2$/dof) and corresponding best-fit bias value ($b$) obtained when fitting our measurements at scales between $1^{\rm o}$ and 20$^{\rm o}$ (for which we have 16 measurements and thus 15 degrees of freedom; numbers in parentheses are the $\chi^2$/dof when we include the six additional measurements at $\theta > 20^{\rm o}$) to our fiducial cosmological model, for 5 cases: 1) No corrections are applied (`No Corrections'); 2) corrections for stars and sky brightness are applied (`Corrections'); 3) $A_{star}$ is subtracted for each star within a pixel of circular area corresponding to the radius $r_{star}$ and a correction for sky background is applied as in case 1 ($-A_{star}$); 4) we iteratively determine weights to apply to the LG density field, as described in Section \ref{sec:csys} (Weights); and 5) we apply the cut $d_{\perp} > 0.5564$ and subtract 0.0034 from each $z_{phot}$ when selecting objects from the south and repeat the $-A_{star}$ procedure ($\Delta$South).} 
\begin{tabular}{lccccc}
\hline
\hline
bin& No Corrections &  Corrections &  $-A_{star}$ & Weights & $\Delta$South \\
\hline
 &$\chi^2$/dof, $b$ & $\chi^2$/dof, $b$ & $\chi^2$/dof, $b$, $r_{star}$ & $\chi^2$/dof, $b$ & $\chi^2$/dof, $b$ \\ 
\hline
1& 0.99 (1.0), 2.16$^{+0.07}_{-0.06}$ & 0.79 (0.74), 2.12$\pm0.07$ & 0.79 (0.75), 2.12$\pm0.07$, 7.56$^{\prime\prime}$ & 0.91 (0.82), 2.08$\pm0.07$ & 0.79 (0.70) 2.11$\pm0.07$\\
2&  3.9 (3.5), 2.26 & 1.8 (1.5), 2.08$\pm0.07$ & 1.8 (1.6), 2.07$\pm0.07$, 10.6$^{\prime\prime}$ & 1.9 (1.7), 2.03$\pm0.07$ & 2.1 (1.8) 2.10$\pm0.07$\\
3& 7.0 (5.8), 2.62 & 0.99 (1.1), 2.20$\pm0.07$ & 1.0 (1.1), 2.21$\pm0.07$, 12.0$^{\prime\prime}$ & 1.1 (0.97), 2.16$\pm0.07$& 0.97 (1.0) 2.23$\pm0.07$\\
4&  6.4 (5.7), 2.63 & 1.0 (1.0), 2.14$\pm0.10$ & 0.97 (0.97), 2.17$^{+0.10}_{-0.09}$, 10.9$^{\prime\prime}$ & 0.64 (0.56), 2.12$^{+0.10}_{-0.09}$ & 0.38 (0.43) 2.10$^{+0.10}_{-0.09}$\\
\hline
\label{tab:bmin}
\end{tabular}
\end{minipage}
\end{table*}

\section{Clustering in Photometric Redshift Bins}
\label{sec:wmeas}
\subsection{Auto- and Cross-correlation Function Measurements}
In order to investigate the photometric redshift catalog, we have measured the angular auto- and cross-correlation functions of galaxies in the four photometric redshift bins summarized in Table \ref{tab:bins}.  This will allow us to investigate how systematics affect particular photometric redshift bins and if the redshift distributions are the same as suggested by the training data.

Fig. \ref{fig:w4pansub} displays the measured auto-correlation functions for LGs, multiplied by $\theta$, in our four photometric redshift bins with no corrections (black triangles), a correction for stars and sky-background (calculated as described in Section \ref{sec:cc}; red squares, `$C_{star,sky}$'), subtracting the effective masked area per star, $A_{star}$, for every star in each pixel and correcting for sky background (blue circles; `$-A_{star}, C_{sky}$'), apply weights based on the $\bar{n}/\bar{n}_{tot}(sys)$ relationships iteratively determined in the order $n_{star}$, airmass, seeing, $A_r$, $d_{\perp}$ offset, sky (open green triangles; `Weights'), and apply the cut $d_{\perp} > 0.5564$ and subtract 0.0034 from each photometric redshift for objects in South while also applying the $-A_{star}, C_{sky}$ correction (open orange squares; `$\Delta$South'). Correcting for either seeing or Galactic extinction makes negligible difference. For each bin, the displayed error-bars are the jack-knife errors. 

Interestingly, only bin 1 and bin 4 show any significant effect from sky background; for bins 1 through 4 the values of $\epsilon_{sky}$ for $C_{star,sky}$ are 0.098, 0.021, 0.034, 0.137. This result is consistent with the assertion that the dependence on sky background is related to its effect on the magnitude errors. The lowest redshift bin should show the largest effect from objects scattering around the $d_{\perp} > 0.55$ cut (as can be inferred from the difference between the red and black curves in Fig. \ref{fig:nznsrat}), and the highest redshift bin should show the largest effect from objects scattering across the faint magnitude limit. 

For each of the measurements displayed in Fig. \ref{fig:w4pansub}, we determine the best-fit bias, $b$, given our fiducial cosmological model. In each panel of Fig. \ref{fig:w4pansub}, the black curve displays the best-fit model when fit to the measurements with the $C_{star,sky}$ corrections applied. We fit to angular scales $1^{\rm o} < \theta < 20^{\rm o}$ (the equivalent physical separation for a $1^{\rm o}$ separation is 23.2$h^{-1}$Mpc at $z=0.5$), for which there are 16 data points in each redshift bin. The best-fit $b$ and the associated $\chi^2$ per degree of freedom ($\chi^2$/dof) are listed for each redshift bin and each of the five separate estimations of $w(\theta)$ in Table \ref{tab:bmin}. Given that we are using a theoretical covariance matrix to compare between our measured and model $w(\theta)$, we have 15 dof regardless of the corrections that we apply. The covariance matrices assume there is no added covariance due to systematics, and the corrections are an attempt to remove this covariance and allow the proper comparison between measurement and model. 

In every case, applying some form of correction reduces the $\chi^2$/dof compared to the case when no corrections are applied. In each redshift bin other than 1, the $\chi^2$/dof is greater than 3 when no corrections are applied. When corrections are applied, the $\chi^2$/dof are all less than 2, and only for bin 2 are they significantly greater than 1. After each of the minimum $\chi^2$/dof reported in Table \ref{tab:bmin}, we list, in parentheses, the $\chi^2$/dof obtained when we fit to a maximum scale of $60^{\rm o}$ (which is the largest scale to which we measure $w[\theta]$) and keep the model the same as the best-fit for $1^{\rm o} < \theta < 20^{\rm o}$. This tests the consistency of the 6 measurements with $20^{\rm o} < \theta < 60^{\rm o}$. Notably, none of the $\chi^2$/dof become significantly worse. 

The most extreme results are obtained when the $\Delta$South corrections are applied. For bin 2, only 0.8\% of $w(\theta)$ measurements consistent with our fiducial model would have a $\chi^2$/dof greater than the value of 2.1 that we find, while for bin 4 we find $\chi^2$/dof = 0.38 and would expect 98.4\% of $w(\theta)$ measurements to have a greater value. As isolated cases, neither result is particularly remarkable. However, regardless of the correction technique, the $\chi^2$/dof for bin 2 are all greater than 1.8. This result is caused, in large part, by the $w(\theta)$ measurements at $\sim2^{\rm o}$, which are considerably smaller than the $w(\theta)$ predicted by our best-fit model.

In order to subtract the effective area per star, $A_{star}$, for every star in each pixel, we use a different value of $r_{star}$ in each photometric redshift bin. We determine these values by fitting for the value of $r_{star}$ that makes $\bar{n}/\bar{n}_{tot}(n_{star})$ most consistent with one for the objects in the redshift bin (just as was done for the full sample as described in Section \ref{sec:fst}). Further, we should expect slightly different values of $r_{star}$ given that Fig \ref{fig:ntararoundstar} shows different relationships for different magnitude LGs, and the average magnitudes are different in each photometric redshift bin. 

The $A_{star}, C_{sky}$ (blue circles) and $C_{star,sky}$ (red squares) corrections result in nearly identical measurements. Thus the best-fit bias values (as presented in Table \ref{tab:bmin}) differ by no more than 1.5\%. Interestingly, the jack-knife errors are smaller in general when we subtract $A_{star}$ --- suggesting that this action reduces the fluctuations within the sample.

When we use the Weight method (open green triangles) to correct our measurements, the resulting $w(\theta)$ amplitudes are slightly smaller than any of the other measurements. This translates to marginally lower (between 1.0\% and 2.5\% compared to the `Corrections'' values) best-fit bias values. This suggests that the Weight method may slightly suppress some true fluctuations. The jack-knife errors in the Weight case are similar to those determined using the $A_{star}, C_{sky}$ correction --- applying the weights to the density field decreases the sample variance. The consistency between the best-fit models and the data are also quite similar to the $A_{star}, C_{sky}$ results, except for bin 4, where the $\chi^2$/dof decreases significantly.

Imposing the $d_{\perp} > 0.5564$ cut and subtracting 0.0034 from $z_{phot}$ for objects in South and applying the $A_{star}, C_{sky}$ correction ($\Delta$South; orange open squares), only makes a significant difference in bin 4. The best-fit bias values change by less than 1\% compared to the $A_{star}, C_{sky}$ best-fit values for every bin other 4, where find a 2.4\% decrease. Again compared to the $A_{star}, C_{sky}$ results, we find a marginal decrease in the $\chi^2$/dof values for bins 1 and 3, but for bin 2, we find a marginal increase. The increase in the $\chi^2$/dof for bin 2 is driven mainly by the $w(\theta)$ measurement at 1.8 degrees. For bin 4, the $\chi^2$/dof decreases by more than 50\%, and as can be seen in the bottom right panel of Fig. \ref{fig:w4pansub}, nearly all of the measurements at scales $> 3^{\rm o}$ become more consistent with the model.

\begin{figure}
\includegraphics[width=84mm]{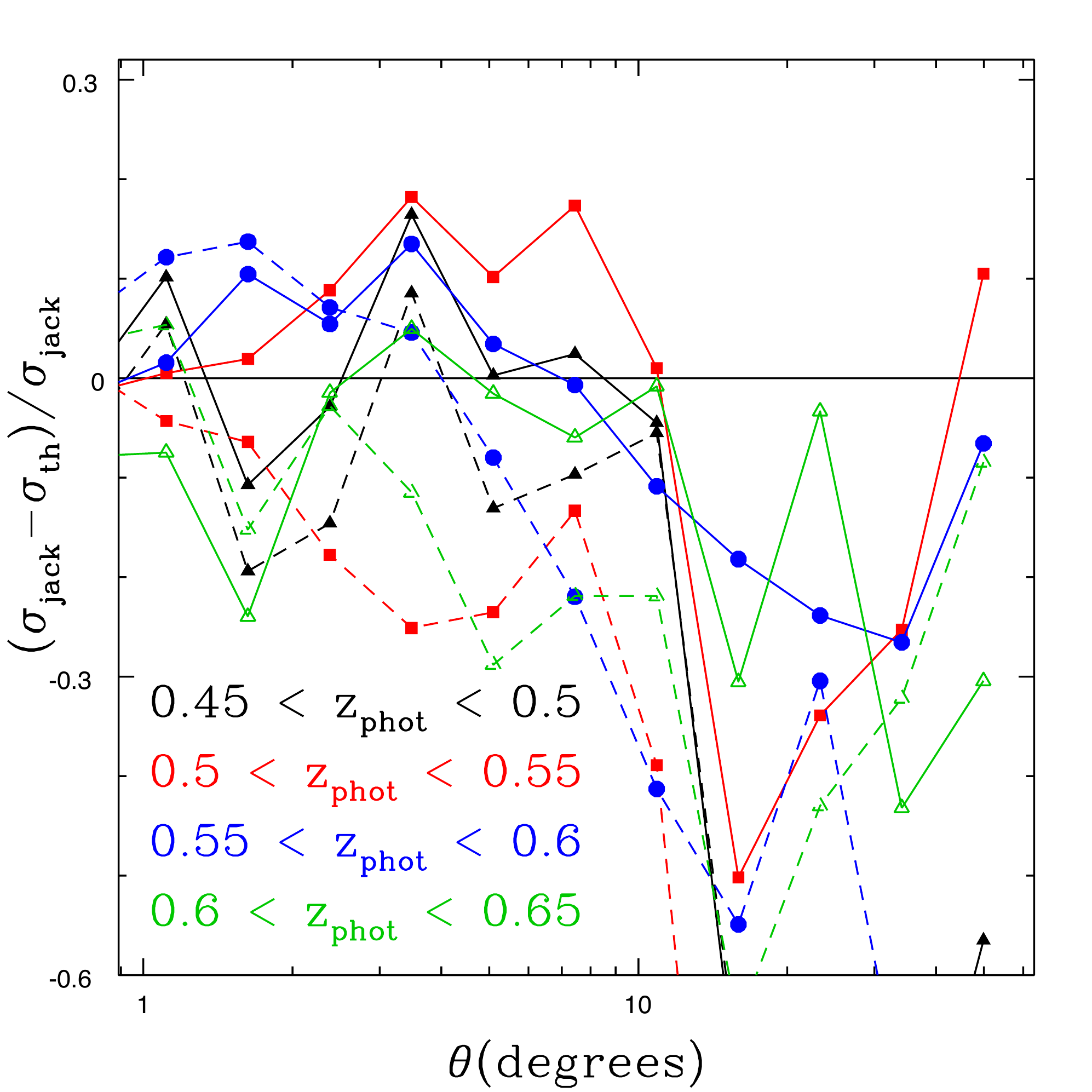}
\caption{The difference between the jack-knife error we estimate, $\sigma_{jack}$ and the theoretical error we calculate, $\sigma_{th}$, divided by $\sigma_{jack}$ for the four redshift bins $0.45 < z_{phot} < 0.5$ (black triangles),  $0.5 < z_{phot} < 0.55$ (red squares), $0.55 < z_{phot} < 0.6$ (blue circles), $0.6 < z_{phot} < 0.65$ (open green triangles). Solid lines connect the points representing the jack-knife errors and covariance matrix when no corrections are made, and the dashed lines connect the points representing the case where $A_{star}$ is subtracted for each star within a pixel of circular area corresponding to the radius $r_{star}$.}
\label{fig:thjackcom}
\end{figure}

In all of the bins, applying some form of correction reduces the $\chi^2$/dof values, and for bins 2 through 4 the corrections change the reduced $\chi^2$ by at least 1.8 (and by as much as 5.4 for bin 4).  Further, the general agreement between the different methods of correction suggest they can all be applied to recover measurements that more closely represent the true clustering of LGs. However, we note that they do not recover the exact same results, suggesting there is some level of systematic error that must be accounted for when similar measurements are used to constrain cosmological parameters. For bins 2 and 3, the variation in the best-fit bias values is similar to the 1$\sigma$ errors --- suggesting the the systematic uncertainties introduced by the need for corrections are a approximately as large as the statistical uncertainties. 

Considering all of the results, we find the bias of the LGs is nearly constant as a function of redshift, with slight evidence of a decrease from high to low redshift. This is close to what we expect for a sample selected to be approximately passively evolving. Such a sample of $b \sim 2$ galaxies will undergo a $\sim 4\%$ decrease in bias over the redshift range $0.475 < z < 0.625$ (see, e.g., \citealt{Fry96,Teg98}). More importantly without any corrections, one might assume our model of large-scale clustering is grossly in error. However, with the corrections, we are given no reason to doubt the standard cosmological model. This is consistent with the results of \cite{CrocceDR7}, whose $w(\theta)$ measurements, at similar redshifts to our own, are consistent with a $\Lambda$CDM model for scales $\theta < 5^{\rm o}$.

Figure \ref{fig:thjackcom} displays the fractional differences we find between the jack-knife errors and the theoretical errors we calculate, for the four photometric redshift bins we use. The solid lines connect the data that represent the case where we apply no corrections, while the dashed lines connect the data representing the $-A_{star}, C_{sky}$ corrections. The results are noisy, but the jack-knife and theoretical errors are similar at scales $< 10^{\rm o}$, while the theoretical errors are larger at greater scales. We note that the agreement is due, in part, to the large differences in the best-fit bias between the cases where corrections are and are not applied. The jack-knife errors are larger when no corrections are applied to the density field, but the best-fit bias is larger as well (see Table \ref{tab:bmin}), and thus the amplitudes of the jack-knife and theoretical errors are similar in both cases. In most cases, the jack-knife errors are smaller than the theoretical estimate at large scales, but this effect is more dramatic when corrections are applied to the density field. One would expect that the jack-knife errors should under-estimate the true uncertainty as the scale grows larger and the different jack-knife regions thus become more correlated. 

\begin{figure*}
\begin{minipage}{7in}
\centering
\includegraphics[width=180mm]{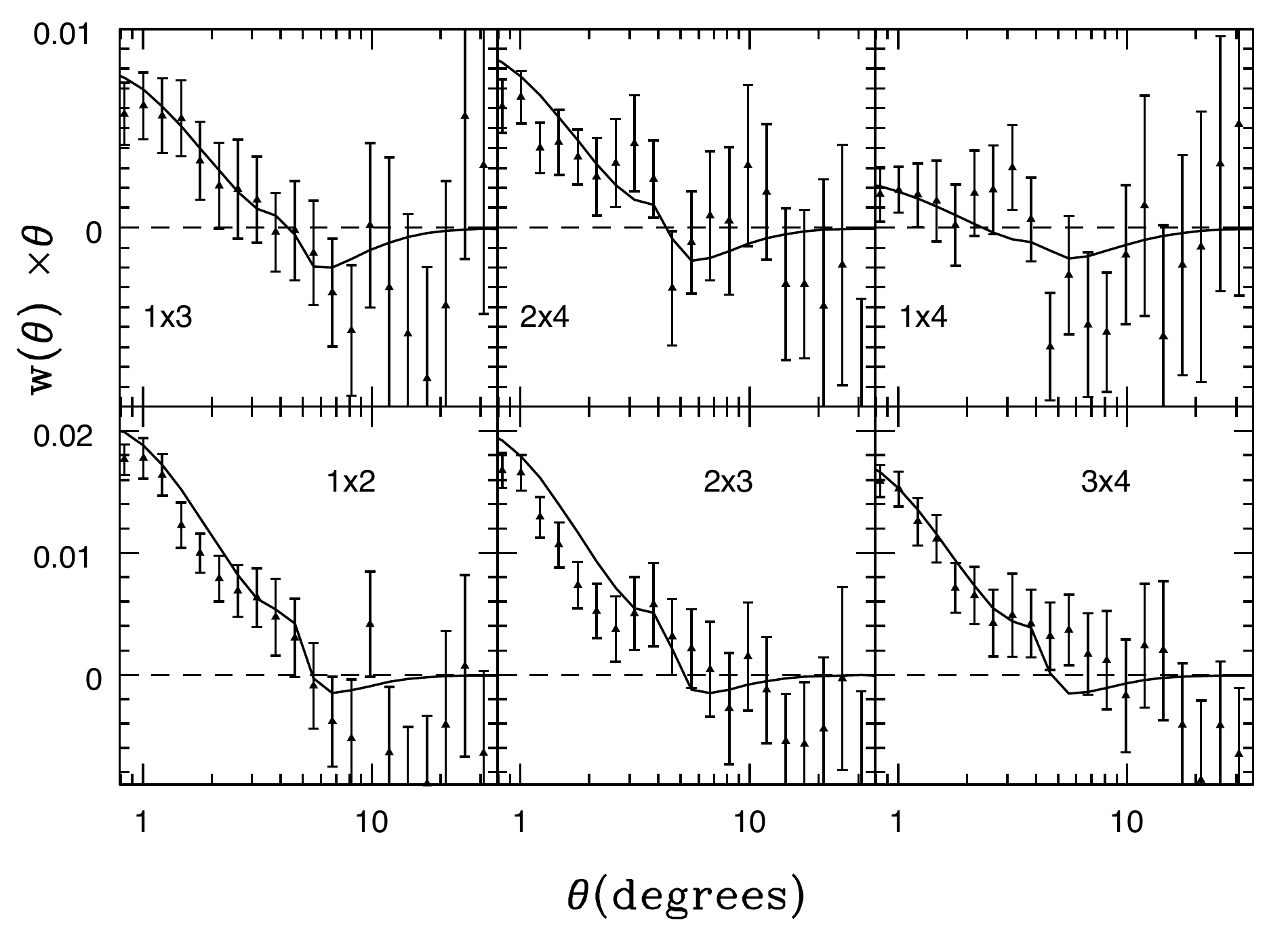}
\caption{The measured angular cross-correlation function measurements between our four photometric redshift bins $0.45 < z_{phot} < 0.5$ (1), $0.5 < z_{phot} < 0.55$ (2), $0.55 < z_{phot} < 0.6$ (3), and $0.6 < z_{phot} < 0.65$ (4), subtracting the systematic effects of stars and sky background via Eqs. \ref{eq:dgds} - \ref{eq:ccc}.  The curves display the theoretical models for $b_1 = 2.08$, $b_2 = 1.96$, $b_3 = 2.16$, and $b_4 = 2.11$, where the bias used is the geometric mean of the bias for the two bins involved.}
\label{fig:wbinall}
\end{minipage}
\end{figure*}

Fig. \ref{fig:wbinall} presents the angular cross-correlation functions between our four photometric redshift samples, after applying corrections for stars and sky background.  For comparison, we display model curves where we assume a bias equal to the geometric mean of the bias of the two bins being cross-correlated (i.e., for 2x3, $b=2.06$). In general, the amplitudes of the cross-correlations are consistent with the redshift distributions determined from our testing set. The cross-correlations that are least consistent are 1x2 and 2x3. The inconsistency could be due to a number of factors. Apart from the redshift distributions being incorrect, it is possible that the bias of the galaxies contributing to the cross-correlation (e.g., for 2x3, the high redshift edge of the $0.5 < z_{phot} < 0.55$ distribution) is lower than for the overall sample. This is likely if objects with lower bias also have larger photometric redshift errors. The facts that the model curves for 1x2 and 2x3 are only marginally outside of the 1$\sigma$ errors and that the agreement appears excellent for the other cross-correlations suggest that there is no significant disagreement. If we do not apply the corrections to the measurements, they are greatly divergent from the models.
\begin{figure*}
\begin{minipage}{7in}
\centering
\includegraphics[width=180mm]{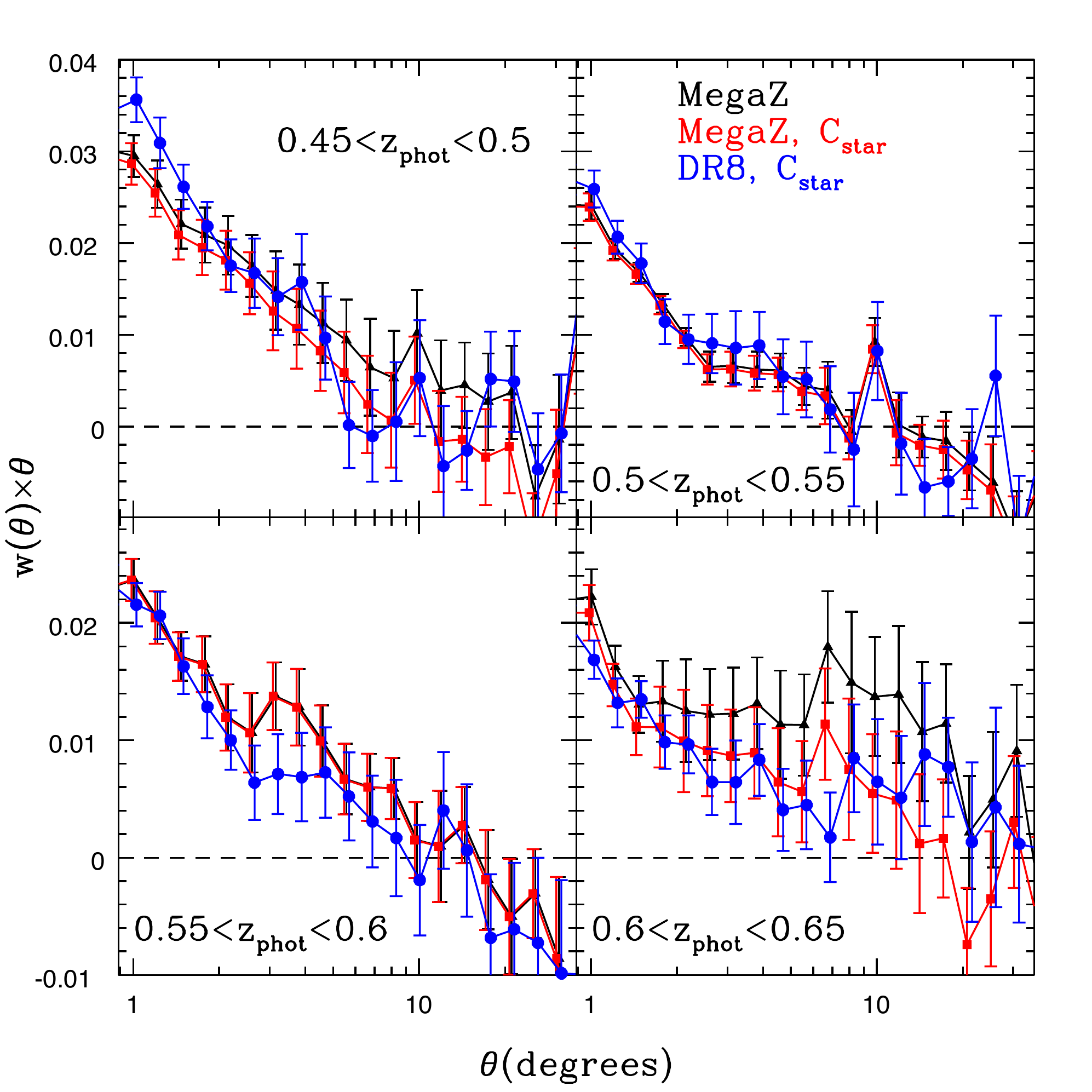}
\caption{The measured angular auto-correlation functions in four photometric redshift bins, $0.45 < z_{phot} < 0.5$ (top-left), $0.5 < z_{phot} < 0.55$ (top-right), $0.55 < z_{phot} < 0.6$ (bottom-left), and $0.6 < z_{phot} < 0.65$ (bottom right). Black triangles display the results using the same catalog (and cuts on it) as Thomas et al. 2010b. Red squares show these MegaZ measurements when they are corrected for stars and the blue circles are the measured $w(\theta)$ using our catalog and correcting for stars, as described in Section \ref{sec:wcorr}.  }
\label{fig:mzwcom}
\end{minipage}
\end{figure*}

\subsection{Comparison With MegaZ}
The MegaZ-LRG DR7 catalog (MegaZ hereafter) is a photometric redshift catalog similar to our own \citep{thomas10}. It used ANNz to train SDSS DR7 objects with similar color selection to ours and spectroscopic redshifts from the 2SLAQ survey \citep{Cannon}. Compared to the sample used in \cite{thomas10}, the most notable differences are that they impose a cut $i_{deV} < 19.8$ while the BOSS survey uses $i_{cmod} < 19.9$ and the sliding cut defined by Eq. \ref{eq:slide}. Further, 2SLAQ spectra were targeted for objects with $i_{fibre} < 21.4$ (our fibre magnitude cut is $i_{fibre2} < 21.7$). 

The MegaZ data and its corresponding mask is publicly available\footnote{http://zuserver2.star.ucl.ac.uk/$\sim$sat/MegaZ/MegaZDR7.tar.gz}. Similar to our catalog, there is a photometric redshift and a probability that the object is a galaxy (we denote this $p_{sgM}$). We employ the same cuts on the MegaZ catalog as \cite{thomas10}, the most notable one being $p_{sgM} > 0.2$. The different target selections result in discrepancies between the overall redshift distributions. These disagreements are most extreme for the lowest and highest redshift bins. Weighting by $p_{sg}$, our catalog has 20\% of its objects with $0.45 < z_{phot} < 0.5$, while  the MegaZ catalog has 36\% of its data in this redshift bin. The $0.6 < z_{phot} < 0.65$ photometric redshift bin contains 14\% of our data, while only 10\% of the MegaZ data have photometric redshifts in this range. Our overall number density is slightly smaller, however, as for $0.45 < z_{phot} < 0.65$, we have a number density of 88 deg$^{-2}$ (872,921 objects over 9,913 deg$^{2}$) and MegaZ has 93 deg$^{-2}$ (723,556 objects over 7746 deg$^{2}$).

The mask provided by MegaZ is in the Healpix format at N$_{\rm side}$$=$1024, and we can therefore calculate $w(\theta)$ using nearly identical methods as described in Section \ref{sec:cc}. We calculate $w(\theta)$ for MegaZ data in the same four photometric redshift bins as our own catalog (we note these are also the photometric redshift bins used by \citealt{thomas10}). We display these measurements, compared to our own, in Fig. \ref{fig:mzwcom}. The black triangles display the measurements we obtain when we cut objects from the MegaZ catalog with $p_{sg} < 0.2$, while the red squares display the results when we correct these same MegaZ measurements for stars in the DR7 area, using the method described in Section \ref{sec:wcorr}. The blue circles display the measurements, for the full DR8 area, when we correct $w(\theta)$ of our catalog for stars.

The small scale amplitudes of the MegaZ measurements and our own are not directly comparable, because the redshift distributions may be different. However, \cite{thomas10} find a significantly smaller bias in their $0.45 < z_{phot} < 0.5$ bin than their other bins, whereas we find only a small difference in the bias of our different photometric redshift bins (see Table 1). It therefore makes sense that the MegaZ amplitudes are significantly smaller than ours. For the middle two redshift bins, the small scale amplitudes of the MegaZ measurements and our own are generally consistent with the displayed jack-knife errors. This result is no surprise given that we would expect the (small) differences in the target selection to only significantly affect the highest and lowest portions of the redshift distribution. Finally, \cite{thomas10} found a substantially higher bias in the MegaZ $0.6 < z_{phot} < 0.65$ sample to their other bins, while for our sample, the bias in this bin is quite similar to that of the $0.55 < z_{phot} < 0.6$ bin. Therefore, the MegaZ amplitudes are larger than ours for $0.6 < z_{phot} < 0.65$.

Correcting for stars makes a substantial difference in the large-scale clustering of the MegaZ data in the lowest and highest redshift bins, but has little effect in the middle two bins. We note that our measurements show no appreciable change when we use only the DR7 area, suggesting that the differences between the two samples must be due to additional systematic effects on either the MegaZ data or our own. 

We note that \cite{thomas10} measured $C_{\ell}$ with these MegaZ samples, and found significant deviations in the expected clustering (from $\Lambda$CDM models) to be confined to the lowest $\ell$ bins. For measurements of $w(\theta)$, the covariance between angular scales implies that an excess at low $\ell$ will affect measurements at smaller angular scales. Given that 
\begin{equation}
  w (\theta) = \sum_{\ell \geq 0} \left(\frac{2\ell+1}{4\pi}\right)
    P_{\ell}(cos\theta)C_{\ell},
   \label{eq:clw}
\end{equation}
where $P_{\ell}$ are Legendre polynomials, one can determine that a 400\% excess for $\ell < 5$ and a 50\% excess for $5 < \ell < 10$ (similar to the excess found in the $0.6 < z_{phot} < 0.65$ bin by \citealt{thomassys}) translate to a 30\% larger $w(\theta)$ at $\theta= 3^{\rm o}$. This estimate is roughly consistent with the difference between the un-corrected (black triangles) and corrected (red squares) MegaZ $w(\theta)$, displayed in the lower-right panel of Fig. \ref{fig:mzwcom} --- i.e., we find the systematic effect of stars of the $0.6 < z_{phot} < 0.65$ MegaZ $w(\theta)$ to be consistent with the low $\ell$ excess found by \cite{thomassys}. 

Failure to apply correction for the systematic effects described in this paper would clearly bias the cosmological parameters one could determine based on our $w(\theta)$ measurements. However, our results suggest that $C_{\ell}$ measurements for $\ell > 10$ might not result in biased measurements. We therefore have no reason to believe that the measurements of cosmological parameters by \cite{thomas10}  have significant systematic errors associated with them. The magnitude of potential systematic errors on the cosmological parameters determined with $C_{\ell}$ spectra is studied in detail in Ho et al. (in prep.).

\section{Conclusions}
\label{sec:con}
We have investigated the systematic effects on the angular distribution and spectroscopic/photometric redshift distributions of objects matching the BOSS CMASS selection. We find that not correcting for the foreground presence of stars, which effectively mask small areas of the sky, produces a systematic error that is (generally) significantly larger than the statistical error at scales greater than 3$^{\rm o}$. The measured $w(\theta)$, after accounting for foreground stars, are generally consistent with $\Lambda$CDM predictions, even at the largest scales we measure, but are grossly inconsistent ($\chi^2$/dof as large as 6.3) when the effects of foreground stars are ignored. Our primary results can be summarized:

\noindent $\bullet$ We select objects from the SDSS DR8 CAS, using the criteria defined for the BOSS CMASS sample, yielding a sample of 1,065,823 objects within our masked footprint which are matched to 112,778 existing BOSS spectra (see Section 2.1). We train ANNz to output a probability that each of these objects is a galaxy (see Section 2.2).

\noindent $\bullet$ Stars occult a small area of the sky, reducing the ability to detect galaxies in their immediate vicinity (see Fig. 3). For our sample, stars with $i$-band magnitude less than 20.3 have an effect out to at least 10$^{\prime\prime}$ (see the top right panel of Fig. 3). 

\noindent $\bullet$ We account for stellar contamination by weighting every object by the probability that it is a galaxy. When doing so, we find a strong, negative correlation between the number density (in deg$^{-2}$) of objects in our sample and stars (see the red squares in the bottom-left panel of Fig. 4 and the black triangles in the bottom panel of Fig. 5), partially explained by our finding that stars effectively mask their local area.

\noindent $\bullet$ We correct for the effect of stars on the local density of galaxies by assuming each star effectively removes constant amount of area. We determine this area as described in Section 4.1. We find that accounting for this area produces a significant change in $w(\theta)$, especially at large scales (see the left-panel of Fig. 7).

\noindent $\bullet$ We test two methods that can be applied in an attempt to correct the systematic errors introduced by any parameter that can be turned into a map on the sky. The ``Correction'' technique, first developed for large-scale structure measurements by Ho et al. (in prep.), is described in Section 3.3. When this method is applied to correct for the presence of stars, we recover nearly identical results as when we account for the effective area of stars (see the left-panel of Fig. 7). The ``Weights'' method is described in Section 4.3. We find that applying it to stars, Galactic extinction, seeing, sky background, airmass, and \cite{Sch10} offsets in $d_{\perp}$ results in $w(\theta)$ measurements that are nearly identical to those we obtain applying the Correction method to stars and sky background (see the right-panel of Fig. 7).

\noindent $\bullet$ We use ANNz to estimate photometric redshifts for every object in our sample. We find that including axis ratios improves the accuracy of the photometric redshift estimates in our testing set (see Section 5.1), and that including Galactic extinction improves their reliability when applied to the full sample (see Section 5.2).

\noindent $\bullet$ We find an asymmetry in the density of objects in the North and South Galactic caps, which is removed (to within 0.1\%) when we account for the 0.0064 difference in $d_{\perp}$ between the North and South discovered by \cite{Sch10b} (see Section 4.4). This offset in $d_{\perp}$ implies that the photometric redshift estimates in the South are biased by 0.0034 compared to the North. When we correct for this bias, we find that the ratio of the number of galaxies in the South to the number in the North is approximately constant and consistent with the ratio of their areas for $0.46 < z_{phot} < 0.65$ (see Fig. 12). Our $w(\theta)$ measurements for the full sample appear similar to a weighted average of the $w(\theta)$ calculated separately for the North and South (see Fig. 8).

\noindent $\bullet$ We divide our photometric redshift catalog into four photometric redshift slices between $0.45 < z_{phot} < 0.65$, as summarized in Table 2. We measure $w(\theta)$ for each slice, applying the various techniques we developed to correct systematic errors (see Fig. 15). We calculate the bias in each sample when using each of the techniques, assuming the same fiducial $\Lambda$CDM model, the results of which are summarized in Table 3. 

\noindent $\bullet$We find that the magnitude of the corrections are larger than the statistical error for $z_{phot} > 0.5$ and $\theta > 3^{\rm o}$ and that applying some form of correction significantly reduces the minimum $\chi^2$/dof. 

\noindent $\bullet$We find scatter in the best-fit bias values that is similar to their 1$\sigma$ uncertainty, suggesting that the systematic error introduced by the need for corrections is of similar magnitude to the statistical uncertainty.

The presence of foreground stars must be accounted for in any study of large-scale clustering --- including the 3D clustering of the BOSS spectroscopic data. Further, similar tests to those presented here will likely be necessary for the radial distribution of BOSS LGs and its impact on the measured clustering.

The results of our study have strong implications for future photometric redshift surveys (such as DES, PanSTARRs, and LSST). We were able to extensively investigate potential sources of systematics because we are determining $z_{phot}$ estimates in the most ideal of cases: our training sample covers a large area, is representative and $\sim$10\% as large as our full catalog. Further, we can include Galactic extinction values in the training because the training sample covers a range representative of our full sample. 

As documented throughout, it is not just the accuracy of the $z_{phot}$ estimate that is important, but the probability that an object is a galaxy as well. Reliable probabilities that an object is a galaxy are crucial for disentangling the systematic and contamination effects of stars on the density field of LGs. Even though our training set is quite large, the relationship shown in Fig. \ref{fig:pgal} is fairly noisy, implying that identifying robust methods of assigning probabilities that objects are galaxies should be a major focus of forthcoming photometric redshift surveys.

Our ideal conditions may not be replicated often in the future (though the photometry should be much better) and many photometric redshifts will be determined via extrapolation of spectral templates. Robust and exhaustive exploration of potential systematics will be difficult under such circumstances, yet extremely important, and their errors are likely to dominate at large scales. It is encouraging that \cite{thomas10} find very similar best-fit cosmological values when they separately use ANNz and various template based methods to determine their photometric redshifts.

Finally, the results we present suggest that the major systematic effects on our photometric redshift catalog of LGs have been identified and can be corrected for, allowing for robust cosmological measurements. \cite{Ho11} show how the same systematics can be accounted for using the angular power spectrum, and present the cosmological constraints obtained using the catalog of galaxies whose creation is described in this paper.

\section*{Acknowledgements}
AJR is grateful to the UK Science and Technology Facilities Council for financial support through the grant ST/I001204/1. AJR would like to thank Daniel Eisenstein, Martin Crocce, and Nacho Sevilla for valuable interactions. WJP acknowledges support from the European Research Council, the Leverhulme Trust, and the UK Science and Technology Facilities Council through the grant ST/I001204/1. We would like to thank the anonymous referee for comments and suggestions that have improved this paper.

Funding for SDSS-III has been provided by the Alfred P. Sloan Foundation, the Participating Institutions, the National Science Foundation, and the U.S. Department of Energy. The SDSS-III web site is http://www.sdss3.org/.

SDSS-III is managed by the Astrophysical Research Consortium for the Participating Institutions of the SDSS-III Collaboration including the University of Arizona, the Brazilian Participation Group, Brookhaven National Laboratory, University of Cambridge, University of Florida, the French Participation Group, the German Participation Group, the Instituto de Astrofisica de Canarias, the Michigan State/Notre Dame/JINA Participation Group, Johns Hopkins University, Lawrence Berkeley National Laboratory, Max Planck Institute for Astrophysics, New Mexico State University, New York University, Ohio State University, Pennsylvania State University, University of Portsmouth, Princeton University, the Spanish Participation Group, University of Tokyo, University of Utah, Vanderbilt University, University of Virginia, University of Washington, and Yale University.

\begin{appendix}
\begin{onecolumn}
\section{Stellar Contamination}
\begin{figure}
\includegraphics[width=84mm]{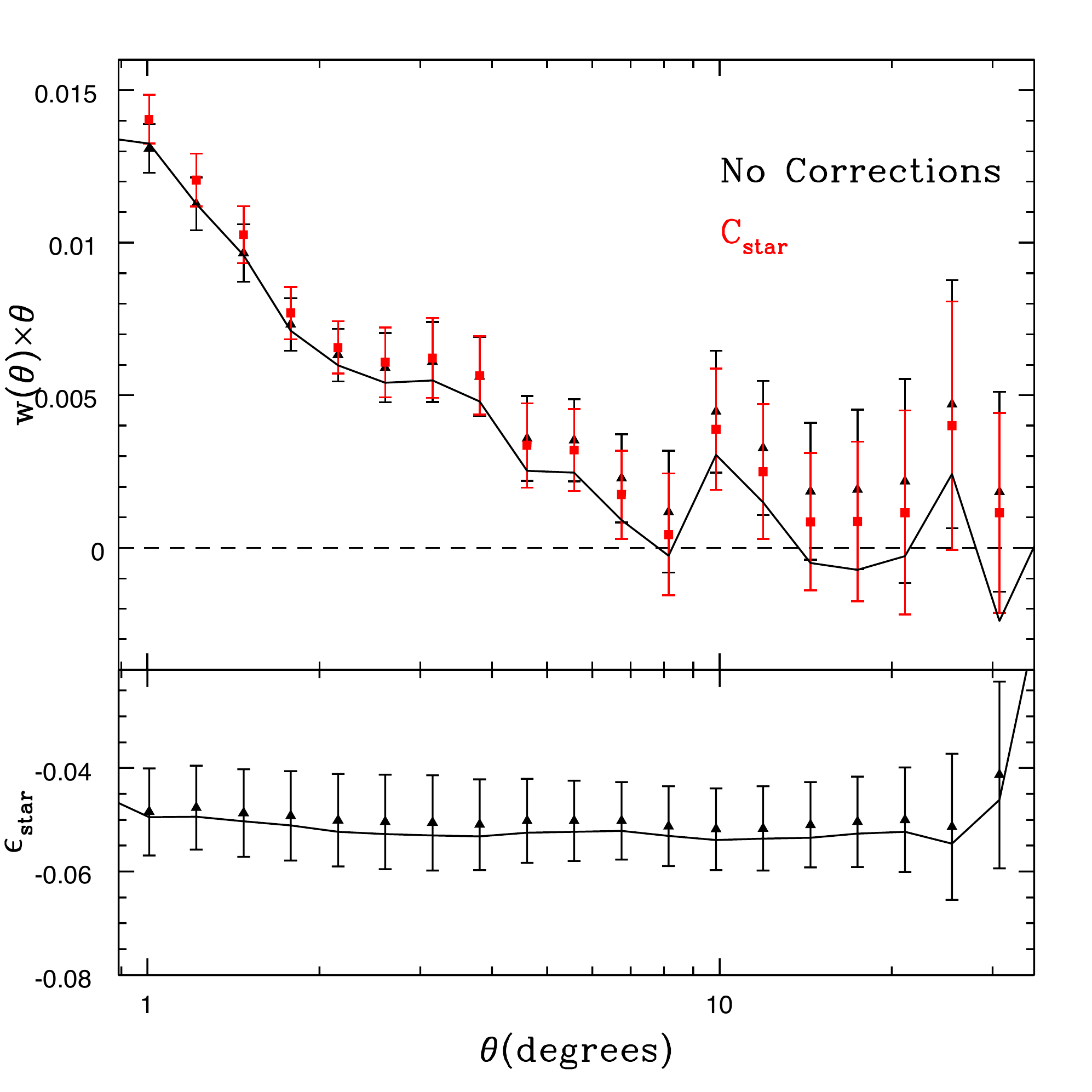}
\caption{Top panel: The measured angular auto-correlation function of all objects in our LG catalog, with equal weighting for every object, multiplied by $\theta$. The red squares display the result when we correct for both the contamination and systematic effects of stars. The black line displays the measurements, corrected for the systematic effect of stars, calculated using the $p_{sg}$ weighting. Bottom panel: The value of $\epsilon$, as a function of $\theta$, based on the cross-correlation function of the un-weighted LG data and the stars. The solid line displays the $\epsilon$ we calculated from the $p_{sg}$ weighted cross-correlation. }
\label{fig:wngs}
\end{figure}

Studies such as \cite{Myers06} have previously described methods for identifying and correcting for stellar contamination. However, Fig. \ref{fig:ntararoundstar} and the left panel of Fig. \ref{fig:ntarall} show that the effect of stars on the density field of LGs is two-fold: 1) some of the objects are stars and these are thus a contaminant and 2) the presence of stars systematically affects the number density of objects. Accounting for both effects makes Eqs. \ref{eq:dgds} - \ref{eq:ccc} more complicated. Given some fraction of the objects that are galaxies, $f_g$, and some fraction that are stars $f_s$ they become:
\begin{equation}
\delta^{\rm o}_g = f_g(\delta^{t}_g+\epsilon\delta_s) + f_s\delta_{sc} ,
\label{eq:dgstar}
\end{equation}

\begin{equation}
w^{t}_g(\theta) = w^{\rm o}_g(\theta)/f^2_g -w_s(\theta)\epsilon^2-w_{sc}(\theta)(f_s/f_g)^2-w_{s,sc}(\theta)2\epsilon f_s/f_g,
\label{eg:w2gt}
\end{equation}
and
\begin{equation}
\epsilon = \frac{w^{\rm o}_{g,s}(\theta)-f_sw_{s,sc}(\theta)}{f_gw_s(\theta)}.
\label{eq:ep}
\end{equation}
where $\delta_{sc}$ is the over-density of stars that act as contaminants and $\delta_s$ is the over-density of all stars. The sum of $p_{sg}$ for the full catalog suggests that 95.9\% of the objects in the catalog are galaxies. Our training data imply that $\sim$ 1.2\% of these objects are quasars. We therefore assume $f_g = 0.96$ and $f_s = 0.03$. We measure $w(\theta)$ and the cross-correlation between LGs and stars using our full catalog, equally weighting each object. If we assume that $\delta_s = \delta_{sc}$ and determine $\epsilon$ vias Eq. \ref{eq:ep}, we find the result plotted in black triangles the bottom panel of Fig. \ref{fig:wngs}. The result is extremely close to the value of $\epsilon$ we obtain when we weight by $p_{sg}$, which is plotted with a solid black line. 

The top panel of Fig. \ref{fig:wngs} displays the $w(\theta)$ measured without any corrections or $p_{sg}$ weights (black triangles). The red squares display the result when we correct for the contamination and systematic effects of stars. This yields similar results as when we correct the $p_{sg}$ weighted measurements for the systematic effect of stars (solid line), but the un-weighted correction is systematically smaller than what would be required for the un-weighted and weighted results to agree. The disagreement is likely due to the fact that we are assuming $\delta_s = \delta_{sc}$ and that we know $f_s$. While we can be confident in the value of $f_g$ by summing $p_{sg}$, the fact that some of the objects are quasars implies we do not have an estimate of $f_s$. Further, it is quite possible that the stars that are mistaken for LGs and are in our sample, have a different distribution than the full distribution of stars. Thus, we are making a number of (possibly incorrect) assumptions.

The approach we adopt in this paper is to weight each LG by its value of $p_{sg}$. Assuming that the $p_{sg}$ values are accurate, this weighting effectively makes $f_g = 1$ (and $f_s = 0$), considerably simplifying the situation. We no longer have to worry about the percentage of objects that are quasars or if the distribution of contaminant stars is different from the full distribution of stars. When there is stellar contamination, the cross-correlation between the observed LG density field and that of the stars will be $\sim f_g\epsilon w_s(\theta) + f_sw_s(\theta)$. The auto-correlation function of stars is positive and their cross-correlation function with LGs is negative (implying $\epsilon$ is negative). Therefore, when there {\rm is} stellar contamination, one is likely to measure a cross-correlation that is $\sim 0$. This would make one (incorrectly!) assume that stars have no effect on the measured auto-correlation function of the LGs. Note that a cross-correlation of $\sim$ 0, when one knows that stellar contamination exists, implies that there {\it must} be a systematic effect of the stars on the density field (since the auto-correlation of the stars is non-zero). We strongly recommend that, given reliable probabilities, one weights an objects by the probability that it is a galaxy when measuring auto-correlation functions.

\section{Systematic Corrections}
\label{app:sysc}
We display the solutions for $\epsilon_i$ based on Eqs. \ref{eq:acc} and \ref{eq:ccc} (as first derived by Ho et al. 2011) for up to three systematics:
\begin{eqnarray}
\epsilon_3 = \frac{w_{g,3}(w_2w_1^2-w_1w_{1,2}^2)-w_{g,1}w_{1,3}(w_2w_1-w_{1,2}^2)+w_1w_{1,3}w_{1,2}w_{g,2}-w_{1,3}w_{g,1}w_{1,2}^2-w_1^2w_{2,3}w_{g,2}+w_1w_{2,3}w_{1,2}w_{g,2}}{w_3w_2w_1^2-w_1w_3w_{1,2}^2+2w_{2,3}w_{1,2}w_{1,3}w_1+w_{1,3}^2w_1w_2-2w_{1,3}^2w_{1,2}^2-w_{2,3}^2w_1^2}\\
\epsilon_2 = \frac{w_1}{w_{1}w_{2}-w_{1,2}^2}\left[w_{g,2}-\frac{w_{1,2}}{w_1}(w_{g,1}-\epsilon_3 w_{2,3})-\epsilon_3 w_{2,3}\right]\\
\epsilon_1 = \frac{1}{w_1}\left[w_{g,1}-\epsilon_2 w_{1,2} -\epsilon_3 w_{1,3}\right]
\end{eqnarray}
\end{onecolumn}

\noindent where $w_{g,i}$ is the cross-correlation function of the galaxies and potential systematic $i$ and $w_{i,j}$ is the cross-correlation function of systematic $i$ and $j$ (and is an auto-correlation function when $i=j$). For only one systematic, the result is simple; one simply subtracts $\epsilon^2w_{sys}(\theta)$ from the measured LG $w(\theta)$. 

We note that the solutions we present are implicitly dependent on $\theta$, but $\epsilon$, as defined, is a constant (Eq. \ref{eq:dgds} does not allow for a $\theta$ dependence). If the measured value of $\epsilon$ changes depending on $\theta$, higher order corrections may be necessary. In this work, we did not find that the error-bars on $\epsilon$ (as displayed in Fig. \ref{fig:wcorr}) warranted applying higher-order corrections. However, the value of $\epsilon$ for the seeing does appear to have a strong dependence on the angular scale, suggesting that if the errors were smaller, higher-order corrections could become necessary.

\end{appendix}

\label{lastpage}
\end{document}